%% file: FlatHolography_updated.tex
\documentclass[a4paper,11pt]{article}
\pdfoutput=1 

\usepackage{jheppub} 
                     
\usepackage[mathscr]{euscript}
\usepackage[T1]{fontenc} 
\usepackage[textsize=footnotesize,textwidth=2.5cm]{todonotes}

\definecolor{mikadoyellow}{rgb} {0.16, 0.254, 0.6}

\newcommand{\rzero}{{\tilde J\over 2\tilde M}}
\newcommand{\tM}{{\tilde M}}
\newcommand{\tJ}{{\tilde J}}
\newcommand{\tu}{{\tilde u}}
\newcommand{\tphi}{{\tilde \phi}}
\newcommand{\tdr}{{\tilde r}}

\usepackage{amssymb,amsmath,latexsym,bm,amsfonts}
\usepackage{graphicx}
\usepackage{longtable}
\usepackage{color,xcolor}
\usepackage{indentfirst}
\usepackage{subfigure}
 \usepackage{comment}
 \usepackage{float}

\input{ShortCommand}

\title{ \boldmath Entanglement Entropy in Flat Holography }

\author[a]{Hongliang Jiang,}
\author[b]{Wei Song}
\author[b]{and Qiang Wen}

\affiliation[a]{Department of Physics, The Hong Kong University of Science and Technology, Clear Water Bay, Kowloon, Hong Kong, P.R.China}
\affiliation[b]{Yau Mathematical Sciences Center,Tsinghua University, Beijing, 100084, China}

\emailAdd{hjiangag@connect.ust.hk}
\emailAdd{wsong@math.tsinghua.edu.cn}
\emailAdd{wenqqq@mail.tsinghua.edu.cn}

\abstract{BMS symmetry, which is the asymptotic symmetry at null infinity of flat spacetime, is an important input for flat holography.
In this paper, we give a holographic calculation of entanglement entropy and R\'{e}nyi entropy in three dimensional Einstein gravity and Topologically Massive Gravity. 
The geometric picture for the entanglement entropy is the length of a spacelike geodesic which is connected to the interval at null infinity by two null geodesics. The spacelike geodesic is the fixed points of replica symmetry, and the null geodesics are along the modular flow.
Our strategy is to first reformulate the Rindler method for calculating entanglement entropy in a general setup, and apply it for BMS invariant field theories, and finally extend the calculation to the bulk. 
}

\begin{document}
\maketitle
\flushbottom

\section{Introduction}
Holography \cite{tHooft:1993dmi,Susskind:1994vu}, which relates a theory containing gravity in higher spacetime dimensions to a quantum field theory in lower dimensions, is believed to be a promising way to understand quantum gravity. 
In particular, holography for asymptotically locally AdS (AlAdS) spacetimes, the so-called AdS/CFT correspondence \cite{Maldacena:1997re,Gubser:1998bc,Witten:1998qj} is one of the most active research fields. 
A priori, it is not clear whether the rich conceptual achievements from AdS/CFT are contingent to AlAdS.
To understand the generality, it is important to extend the great success of holography beyond the context of AdS/CFT. 
Progresses for non-AlAdS holography includes dS/CFT correspondence \cite{Strominger:2001pn,Anninos:2011ui}, the Shr$\ddot{\text{o}}$dinger or Lifshitz spacetime/non-relativistic field theory duality \cite{Son:2008ye,Balasubramanian:2008dm,Kachru:2008yh,Taylor:2015glc}, the Kerr/CFT correspondence \cite{Guica:2008mu,Bredberg:2009pv,Castro:2010fd,Bredberg:2011hp,Compere:2012jk}, the WAdS/CFT \cite{Anninos:2008fx} or WAdS/WCFT \cite{Detournay:2012pc} correspondence, and illuminating results toward flat holography in four dimensions \cite{Bondi:1962px,Sachs:1962wk,Strominger:2013jfa,Hawking:2016msc} and three dimensions \cite{Barnich:2010eb,Bagchi:2010eg,Bagchi:2012cy,Bagchi:2014iea}.  

A general item in the dictionary of holography is that the asymptotic symmetry for the gravitational theory in the bulk agrees with the symmetry of the dual field theory, if the later exists. For four dimensional flat spacetime,  the asymptotic symmetry group at null future (past) is the BMS${^\pm}$ group, first studied by Bondi, van der Burg, Metzner and Sachs \cite{Bondi:1962px,Sachs:1962wk}. 
In a recent resurgence, 
Strominger \cite{Strominger:2013jfa} pointed out that the diagonal elements of BMS$^+\times$ BMS$^-\rightarrow$ BMS is the symmetry of the $S-$matrix. 
BMS group is connected to infra properties of scattering amplitude \cite{He:2014laa,Cachazo:2014fwa}, and memory effects \cite{Zeldovich}, in a triangle \cite{Strominger:2014pwa}. See the lecture notes \cite{Strominger:2017zoo} for a  review. As a simple toy model, the three dimensional version of BMS group has also generated lots of interested.
 BMS$^\pm_3$ on $\mathcal I^\pm$ was discussed in \cite{Ashtekar:1996cd,Barnich:2006av,Barnich:2010eb,Bagchi:2012yk,Detournay:2014fva}. Interesting developments include
connections with Virasoro algebra \cite{Barnich:2006av}, isomorphism between BMS algebra and Galileo conformal algebra \cite{Bagchi:2010eg}, representations and bootstrap \cite{Barnich:2012aw,Barnich:2012rz,Barnich:2014kra,Barnich:2015uva,Campoleoni:2016vsh,Oblak:2016eij,Carlip:2016lnw,Bagchi:2016geg,Batlle:2017llu}. 
Flat holography based on BMS$_3$ symmetry was proposed in \cite{Bagchi:2010eg,Bagchi:2012cy} and supporting evidence can be found in \cite{Bagchi:2012xr,Barnich:2012xq,Bagchi:2016bcd}. The antipodal identification in three dimensions was discussed in  
\cite{Prohazka:2017equ,Compere:2017knf}.   

One useful probe of holography is the entanglement entropy, which describes the correlation structure of a quantum system. In the context of AdS/CFT correspondence, Ryu and Takayanagi \cite{Ryu:2006bv,Ryu:2006ef} (RT) proposed  that the entanglement entropy is given by the area of a codimension-two minimal surface in the bulk, which anchored on the entangling surface of the subsystem on the boundary. A covariant version was proposed by Hubeny, Rangamani and Takayanagi (HRT) \cite{Hubeny:2007xt}. Using AdS/CFT for Einstein gravity, RT and HRT proposal have been proved by \cite{Casini:2011kv,Hartman:2013mia,Faulkner:2013yia,Lewkowycz:2013nqa, Dong:2016hjy}.
It is interesting to ask if the connection between spacetime structure in the bulk and entanglement in the boundary still exist beyond the context of AdS/CFT, and if so, how it works for non-AlAdS spacetimes. So far, in the literature, there are three approaches. 
The first approach is to start with the RT or HRT proposal, and study the implications in the 
holographic dual, see \cite{Li:2010dr,Anninos:2013nja,Basanisi:2016hsh,Gentle:2015cfp,Sun:2016dch}.  The second approach is to directly propose a prescription in the bulk, and check its consistency \cite{Sanches:2016sxy,Bakhmatov:2017ihw}.   
The third approach, which we will advocated in the current paper, is to derive an analog of RT proposal  using the dictionary of holography, along the lines of \cite{Casini:2011kv,Hartman:2013mia,Faulkner:2013yia,Lewkowycz:2013nqa, Dong:2016hjy}. In \cite{Song:2016pwx,Song:2016gtd},  holographic entanglement entropy in Warped AdS$_{3}$ spacetime was derived by generalizing the  gravitational entropy \cite{Lewkowycz:2013nqa} and Rindler method \cite{Casini:2011kv,  Castro:2015csg}, respectively.  Interestingly, it was found that the HRT proposal indeed need to be modified, and moreover the modification depends on different choices of the boundary conditions which determines  the asymptotic symmetry group. Another important lesson is that the Rindler method \cite{Casini:2011kv}, which maps entanglement entropy to thermal entropy by symmetry transformations, can be generalized to non-AlAdS dualities. 

In the context of flat holography, entanglement entropy for   field theory with BMS$_3$ symmetry was considered in \cite{Bagchi:2014iea} using  twist operators.  Using the Chern-Simons formalism of 3D gravity,  \cite{Basu:2015evh} took the Wilson line approach \cite{Ammon:2013hba,deBoer:2013vca,Castro:2014tta}, and found agreement with  \cite{Bagchi:2014iea}. However, a direct calculation in metric formalism is still missing. No geometric picture has been proposed and  it is not clear
 whether RT (HRT) proposal is applicable for asymptotic flat spacetime.  
 In this paper, we will address this question along the lines of \cite{Song:2016pwx,Song:2016gtd}.

In this paper, the Rindler method is formulated in general terms, without referring to any particular example of holographic duality.  
We argue that the entanglement entropy for a subregion $\mathcal A$ is given by the thermal entropy on $\tilde B$, if there exists a Rindler transformation from the causal development of $\mathcal A$ to $\tilde B$.  Moreover, under such circumstances, the modular Hamiltonian implements a geometric flow generated by the boost vector $k_t$. Then we apply this generalized Rindler method to holographic dualities governed by BMS symmetry, and provide a holographic calculation of entanglement entropy and R\'{e}nyi entropy in Einstein gravity and Topologically massive gravity. 
On the field theory side, our result of entanglement entropy agrees with that of \cite{Bagchi:2014iea} obtained using twist operators \cite{Calabrese:2004eu}.
On the gravity side, by extending the Rinlder method to the bulk,  we provide a holographic calculation of the entanglement entropy in metric formalism and provide a geometric picture (see figure \ref{Penrose}). We also expect  a generalization in higher dimensions.   

The geometric picture (figure \ref{Penrose}) for holographic entanglement entropy in three dimensional flat spacetime involves three special curves, a  spacelike geodesic $\gamma$, and two  null geodesics $\gamma_{\pm}$.  $\gamma$ is the set of fixed points of the bulk extended modular flow $k_t^{\text{bulk}}$ (and also fixed points of bulk extended replica symmetry), while $\gamma_\pm$ are the orbits of the boundaey end points $\partial\mathcal{A}_{1,2}$ under $k_t^{\text{bulk}}$. The end points of $\gamma$ is connected to the boundary end points $\partial\mathcal{A}$ by $\gamma_\pm$. The holographic entanglement entropy for the boundary interval $\mathcal{A}$ is given by $\frac{\text{Length}(\gamma)}{4G}$. The main difference between our picture and the RT (HRT) proposal is that the spacelike geodesic $\gamma$ is not directly connected to the boundary end points $\partial\mathcal{A}$.  The reason is that points on $\partial\mathcal{A}$ are the fixed points of the boundary modular flow $k_t$  but not the fixed points of the bulk modular flow $k_t^{\text{bulk}}$.   
Our results are consistent with all previous field theory \cite{Bagchi:2014iea} and as well as a Chern-Simons calculation \cite{Basu:2015evh}. 
In this paper, explicit calculations are done in Bondi gauge at future null infinity. Similar results follows for the past null infinity. We also expect the techniques and results here can be reinterpreted in the hyperbolic slicing \cite{deBoer:2003vf,Ashtekar,Beig,deHaro:2000wj,Compere:2017knf}.

The paper is organized as follows. In section~\ref{rindler} we explain the generalized Rinder method and provide a formal justification for its validity. In section~\ref{BMSGCA}, BMS$_3$ and asymptotically flat spacetimes is reviewed. In section~\ref{FieldEE}, we calculate the EE in the BMS$_3$ invariant field theory (BMSFT) by Rindler method. Then we calculate the entanglement entropy holographically for Einstein gravity in section~\ref{GravityHEE}. In section~\ref{geometry}, we give geometric picture of holographic entanglement entropy. In section~\ref{FlatAdS}, we calculate the HEE by taking the flat limit of AdS. In section \ref{TMG} we calculate the holographic entanglement entropy in topological massive gravity. In section \ref{Renyientropy}, we calculate the R\'{e}nyi  entropy both on the field theory side and the gravity side. In appendix \ref{CardyThermal}, we rederive the ``Cardy formula'' for BMSFT using the BMS symmetries. In appdneix \ref{KillingVec}, we present the Killing vectors of 3D bulk flat spacetime. In appendix \ref{RindlerTsFSC} we give the details of  Rindler coordinate transformations for  BMSFT in finite temperature and on cylinder.

\section{Generalized Rindler method} \label{rindler}
The Rindler method was developed in the context of AdS/CFT \cite{Casini:2011kv} with the attempt to derive the Ryu-Takayanagi formula.
For spherical entangling surfaces on the CFT vacuum, the entanglement entropy can be calculated as follows.
In the  CFT side, certain conformal transformations map the entanglement entropy in the vacuum to the thermodynamic entropy on a Rindler or hyperbolic spacetime. In the bulk, certain coordinate transformations map vacuum AdS to black holes with a hyperbolic horizon.
Using the AdS/CFT dictionary, the Bekenstein-Hawking entropy calculates the thermal entropy on the hyperbolic spacetime, and hence provides a holographic calculation of the entanglement entropy. Going back to vacuum AdS, the image of the hyperbolic horizon then becomes an extremal surface ending on the entangling surface at the boundary.
Recently, the Rindler method has been generalized to holographic dualities beyond AdS/CFT.
The field theory story was generalized to Warped Conformal Field Theories (WCFT) in \cite{Castro:2015csg}, while
the gravity story was generalized to Warped Anti-de Sitter spacetimes (WAdS) in \cite{Song:2016gtd}.  The results are consistent with the WAdS/WCFT correspondence \cite{Anninos:2008fx}.

In this section, we summarized the Rindler method for holographic entanglement entropy, without referring to the details of the holographic pair. The goal is to provide a general prescription which could be potentially used in a broader context. 
Schematic prescriptions in the field theory side and the gravity side are as follows.

\subsection{Field theory calculation of entanglement entropy}


\subsubsection{Generalized Rindler method}\label{Rindlertransformations}
In the field theory, the key step is to find a Rindler transformation, a symmetry transformation which maps the calculation of entanglement entropy to thermal entropy.
Consider a QFT on a manifold $\mathcal{B}$ with a symmetry group $G$, which act both on the coordinates and on the fields. 
The vacuum preserve the maximal subset of the symmetry, whose generators are denoted by ${h_j}$. Consider the entanglement entropy for a subregion $\mathcal{A}$ with a co-dimension two boundary $\p\mathcal{A}$.
Acting on positions, a Rindler transformation $R$ is a symmetry transformation.
The image of $R$ is a manifold $\tilde{\mathcal{B}}$\footnote{ Throughout this paper, we always use tilded variables to describe the $\tilde{\mathcal{B}}$ spacetimes and their bulk extensions after the Rindler transformation. }, and the domain is $\mathcal{D}\subset\mathcal{B}$, with $\mathcal{A}\subset\mathcal{D}$, and $\p\mathcal{A}\subset\p\mathcal{D}$. For theories with Lorentz invariance, $\mathcal{D}$ is just the causal development of $\mathcal{A}$. The image of $\p \mathcal D$ should also be the boundary of $\tilde {\mathcal B}$.
A Rindler transformation $R$ is supposed to have the following features:
\begin{enumerate}
\item The transformation $\tilde{x}=f(x)$ should be in the form of a symmetry transformation. 
\item  The coordinate transformation $x\rightarrow \tilde x$ should  be invariant under some imaginary identification of the new coordinates $\tilde{x}^i \sim \tilde{x}^i+i {\tilde{\beta}}^i$.  Such an identification will be referred to as a ``thermal'' identification hereafter.
\item  
The vectors $\p_{\tilde{x}^i}$ annihilate the vacuum. i.e. 
\be \p_{\tilde{x}^i}=\sum_j b_{ij} h_j\label{vacuum}\,,\ee where $b_{ij}$ are arbitrary constants.
\item Let $k_t\equiv {\tilde \beta}^i \p_{\tilde{x}^i}$, then $k_t$ generates a translation along the thermal circle, and induce the flow 
$\tilde{x}^i(s)=\tilde{x}^i+{\tilde{\beta}}^is$. A thermal identification can be expressed as $\tilde{x}^i\sim \tilde{x}^i(i)$. 
The boundary of the causal domain  $\p \mathcal D$ should be left invariant under the flow. In particular, $k_t$ can only become degenerate\footnote{For Warped conformal field theory \cite{Song:2016gtd}, $k_t$ keeps $\p \mathcal{D}$ invariant, but will not degenerate anywhere for $\alpha\neq0$. } at the entangling surface $\p \mathcal A$, \be {\tilde \beta}^i \p_{\tilde{x}^i}|_{\p \mathcal A}=0\,. \label{fixed}\ee
\end{enumerate}
Now we argue that the vacuum entanglement entropy on $\mathcal A$ is given by the thermal entropy on $\mathcal{ \tilde B}$, if such a Rindler transformation with the above properties can be found. 
Property 2 defines a thermal equilibrium on $\mathcal{ \tilde B}$.
Property 3 implies that the vacuum state on $\mathcal B$ is mapped to a state invariant under translations of ${\tilde x}^i$, which is just the thermal equilibrium on $\mathcal {\tilde B}$. 
The Modular Hamiltonian on $\tilde{\mathcal{B}}$, denoted by $H_{\tilde{\mathcal{B}}}$ then implements the geometric flow along $k_t$.
With property 1,  the symmetry transformation $R$ acts on  the operators by an unitary transformation $U_R$.  By reversing the Rindler map, a local operator $H_{\mathcal{D}}$ on $\mathcal{D}$ can be defined by
 $U_{\mathcal{D}}(s)= U_R \,\tilde U(s) \,U_R^{-1}$, where ${\tilde U}(s)\equiv e^{-i \tilde H s}$ .
Note that $U_{\mathcal{D}}(s)$ generates the geometric flow, and $H_{\mathcal D}\equiv \ln U_{\mathcal{D}}(i)$ is just the conserved charge $Q_{k_t}$ up to an additive constant.
Property 4 indicates that $U_{\mathcal{D}}(s)$ implements a symmetry transformation which keeps $\mathcal D$ invariant. Then we can always decompose the Hilbert space of $\mathcal A$ in terms of eigenvalues of $\mathcal H_{\mathcal D}$. Therefore, the modular Hamiltonian on $\mathcal D$ can indeed be written as $H_{\mathcal D}$, which again  generates  the geometric flow along $k_t$. In particular, the density matrix are related by a unitary transformation \be \rho_{\mathcal A}=U_R \,\rho_{\mathcal{ \tilde B}} \,U_R^{-1}\,.\ee 
Since unitary transformations does not change entropy, the entanglement entropy on $\mathcal A$ is given by the thermal entropy on $\mathcal{ \tilde B}$. 


More explicitly, at the thermal equilibrium, the partition function and density of matrix on $\mathcal{\tilde B}$ can be written as \footnote{If there are internal symmetries, the partition function should be modified accordingly. } \be Z({\mathcal{\tilde B}})=\Tr e^{-\tilde{\beta}^i Q_{\tilde x^i} },\quad \rho_{\mathcal{\tilde B}}\equiv \Tr\, e^{-\tilde{H}}=Z({\mathcal{\tilde B}})^{-1}e^{-\tilde{\beta}^i Q_{\tilde x^i} }\,,\ee where $Q_{\tilde x^i}$ are the conserved charges associated with the translation symmetries. The modular flow is now local, and is generated by $k_t\equiv {\tilde \beta}^i \p_{\tilde{x}^i}$. The action of the flow on positions and fields are given by \bea \tilde{x}^i(s)&=&\tilde{x}^i+{\tilde \beta}^i s\,,\label{flow}\\
\mathcal{\tilde O}(s)&=&  \tilde U(s) \mathcal{\tilde O} \tilde U(-s)\,.
\eea 
The entanglement entropy which is equivalent to thermal entropy is then given by \be S_{\text{EE}}(\mathcal{A})=S(\tilde {\mathcal B})=(1-\tilde{\beta}^i\p_{{\tilde \beta}^i}) Z({\mathcal{\tilde B}})\,.\ee
R\'{e}yni entropy and the Modular entropy \cite{Dong:2016fnf} can be calculated in a similar fashion. 
In fact, $\tau=\sum_i {{\tilde x}^i\over 2\pi\beta_i}$ parameterizes the Rindler time. The thermal identification is just $ \tau\sim \tau+2\pi i$.
The replica trick can be performed by making multiple copies of $\mathcal B$ and impose the periodicity boundary condition 
\bea &&\tau\sim \tau+2q\pi i\\
&&\phi_b(\tau+2 \pi   i)=\phi_b(\tau)
\eea

To actually find the Rindler map, the strategy is to follow the steps below, 
\begin{itemize}
\item Take an arbitrary symmetry transformation, and impose the condition (\ref{vacuum}). This will give a system of differential equations, whose solution will depends on the constants $b_{ij}$. 
\item The temperatures can be read off from the transformations, and will be determined by certain combinations of $b_{ij}$.
\item Further solving condition (\ref{fixed}) will relate the $b_{ij}$ to position and size of the entangling surface $\p\mathcal{A}$.
\end{itemize} 

\subsubsection{``Cardy'' formula }\label{Cardy}
In conformal field theory, $S$-transformation is used to estimate the resulting thermal entropy, leading to an analog of the Cardy formula \cite{Cardy:1986ie,Hartman:2014oaa}.   A successful generalization  has been applied to WCFT in \cite{Detournay:2012pc, Castro:2015csg}, and BMSFT in \cite{Bagchi:2012xr,Barnich:2012xq}. More generally, $S$-transformation can be realized as a coordinate transformation compatible with the symmetry, which effectively switches the spatial circle and thermal circle. In some region of parameters, the partition function is dominated by the vacuum contribution, and the entropy can hence be estimated. 

For our purpose in BMSFT, we revisit the Cary-like formula in appendix \ref{CardyThermal} and obtain the approximated entropy formula for BMSFT on arbitrary torus. Our derivation is based on the BMS symmetries only, without resorting to flat limit of CFT.

\subsection{Holographic entanglement entropy}
The gravity story is the extension of the field theory story using holography. 
There are two possible routes. \begin{itemize}
\item
The first route is to find the classical solution in the bulk which is dual to thermal states on $\tilde{\mathcal{B}}$. 
This can be obtained by extending the boundary coordinate transformation $\tilde{x}=f(x)$ to the bulk, by performing a quotient. More detailed discussion can be found in \cite{Song:2016gtd}.
\item 
The second route is to extend replica symmetry to the bulk along the lines of \cite{Lewkowycz:2013nqa,Dong:2016hjy}. 
The field theory generator $k_t$ has a bulk extension $k_t^{\text{bulk}}$ via the holographic dictionary. Since $\p A$ is the fixed point of $k_t$, we expect a special bulk surface $ \gamma$ satisfying 
\be k_t^{\text{bulk}}|_{\gamma}=0\,,\ee Such a bulk surface will be the analog of RT( HRT ) surface.  
However, if  $k_t^{\text{bulk}}|_{\p A}\neq 0,$ the homologous condition can not be imposed directly. As we will see later, $\gamma$ is connected to $\p A$ by two null geodesics $\gamma_\pm$, which are along the bulk modular flow $k_t^{\text{bulk}}$.
We will discuss a local version of this approach  in a future work \cite{JSW}.
\end{itemize}

\section{Review of BMS group and asymptotically flat spacetime}\label{BMSGCA}
\subsection{BMS invariant field theory}\label{BMSFT}
In this subsection, we review a few properties of  two dimensional  field theory with BMS$_3$ symmetry( BMSFT). On the plane \cite{Bagchi:2009pe},  BMS symmetries are generated by the following vectors 
\bea\label{GCAgenerator}
L_n&=&-z^{n+1}\partial_z-(n+1)z^n w \partial_w \,,\quad\\
M_n&=&z^{n+1}\partial_w\,.
\eea
The finite BMS transformations can be written as  \cite{Barnich:2012xq}
\besub\label{bmsfinite}
\bea
\tilde{z}&=&f(z)\,,\\
\tilde{w}&=&f'(z)w+g(z)  \,.
\eea
\eesub
Let $J(z)$ denote the current associated to the reparameterization of $z$, and let $P(z)$ denotes the current associated to the $z$-dependent shift of $w$.
We can define charges \footnote{We believe the analytic continuations of $z,w$ to complex numbers are inessential. } \cite{Barnich:2012xq}
\bea
\mathcal{L}_n&=& \frac{1}{2\pi i}\oint_{|z|=1} dz z^{n+1} J(z)\,,\\
\mathcal{M}_n&=& \frac{1}{2\pi i}\oint_{|z|=1} dz z^{n+1} P(z)\,.
\eea
The conserved charges satisfy the central extended algebra
\besub\label{GCAcentral}
\beqn
\big[ \mathcal L_n,\mathcal L_m]  &=&  (n-m)\mathcal L_{n+m}+\frac{c_{\text{L}}}{12}(n^3-n)\delta_{m+n,0}\,,\\
\big[\mathcal L_n,\mathcal M_m]  &=&  (n-m)\mathcal  M_{n+m}+\frac{c_{\text{M}}}{12}(n^3-n)\delta_{m+n,0}\,,\\
\big[\mathcal  M_n,\mathcal M_m]  &=&  0\,,
\eeqn
\eesub
where $c_{\text{L}},c_{\text{M}} $ are the central charges.
Under the transformation (\ref{bmsfinite}), the currents transform as
\beqn
 \tilde{   P}(\tilde{z}) &=& (\tilde f')^{2}\,    P(z) +  {c_{\text{M}}\over12}\{\tilde f, \tilde z \},   \\
\tilde{  J}({\tilde z}) &= & (\tilde f')^{2} \,     J(z)  + 2  \tilde  f'\,   \tilde  g'  \,   P(z) +   (\tilde f')^{2}\,  \tilde g\,   P'(z) + {c_{\text{L}}\over12}  \{\tilde f, \tilde z \} + {c_{\text{M}}\over12} [\![  ( \tilde f , \tilde g  ),  \tilde z ]\!],\label{chargetr}
\eeqn
where   $z = \tilde f(\tilde z)$ and $w = \tilde w \tilde f'(\tilde z) + \tilde g(\tilde z)$  is the inverse coordinate transformation of eq.~\eqref{bmsfinite}.  
The ordinary  Schwarzian above is  
\be
\{ \tilde f,\tilde  z  \}=\frac{ \tilde  f''' }{\tilde  f' }-\frac{3}{2} \Big( \frac{\tilde f'' }{\tilde  f' } \Big)^2\,,
\ee
while the ``BMS  Schwarzian''  is given by 
\be
[\![  (  \tilde f , \tilde g ),  \tilde z ]\!]  =\frac{ \big[3 (\tilde f'')^{2}-\tilde f' \tilde f'''    \big]\tilde g'  -3 \tilde f' \tilde f'' \tilde g''+(\tilde f')^2 \tilde g'''
}{(\tilde f' )^3  } .
\ee

In particular, the transformation below maps a plane to a cylinder,
\begin{align}
z= e^{i\phi}\,,\qquad
w= i e^{i\phi} u\,,
\end{align}
 
The currents transform as 
\be
J^{\text{Cyl}}(\phi)=-z^2 J(z)+\frac{c_{\text{L}}}{24}\,,
\qquad P^{\text{Cyl}}(\phi)=-z^2 P(z)+\frac{c_{\text{M}}}{24}\,,
\ee
and
\be\label{PlanCylLM}
\mathcal{L}^{\text{Cyl}}_n  =  \mathcal{L} _n  -  \frac{c_{\text{L}}}{24} \delta_{n,0}\,,\qquad
\mathcal{M}^{\text{Cyl}}_n  =  \mathcal{M} _n  -  \frac{c_{\text{M}}}{24}\delta_{n,0}\,,
\ee
where the generators on the cylinder are defined as
\bea
\mathcal{L}^{\text{Cyl}}_n&=&- \frac{1}{2\pi }\int_0^{2\pi} d\phi e^{in\phi}      J^{\text{Cyl}}(z)\,,\\
\mathcal{M}^{\text{Cyl}}_n&=& - \frac{1}{2\pi }  \int_0^{2\pi} d\phi e^{in\phi}    P^{\text{Cyl}}(z)\,.
\eea

\subsection{BMS as asymptotic symmetry group}\label{flatholography}

Minkowski spacetime is the playground of modern quantum field where the Poincar\'{e} symmetry gives very stringent constraint on the properties of particles. The Poincar\'{e} algebra includes translation and Lorentz transformation, while the Lorentz transformations further consist of spatial rotation and boost.  At the null infinity of flat spacetime, the finite dimensional Poincar\'{e} isometry group is enhanced to infinitely dimensional asymptotic symmetry group,  called BMS group \cite{Bondi:1962px,Sachs:1962wk}.

In three spacetime dimensions, the topology of the boundary at null infinity is  $S^1\times \mathbb{R}$ where $ \mathbb{R}$ is the null direction.
Under proper boundary conditions \cite{Barnich:2006av,Barnich:2010eb}, the general solution to Einstein equation in the Bondi gauge is 
\be\label{SolEOM}
ds^2=\Theta(\phi) du^2-2 dudr+2\Big[ \Xi(\phi) + \frac{u}{2} \partial_\phi \Theta(\phi)\Big] du d\phi +r^2 d\phi^2\,,
\ee where the null infinity is at $r\rightarrow \infty$. 
The asymptotic symmetry group is the three dimensional BMS group, whose algebra is given in the previous subsection.

BMS$_3$ group is generated by the super-translation and super-rotation. The super-translation can be thought as the translation along the null direction which may vary from one point to another in   $S^1$, while the super-rotation is the diffeomorphism of  $S^1$.
Furthermore, the corresponding conserved charges generates BMS$_3$ group on the phase space. The infinitely dimensional algebra now has central extensions, which also coincide with \eqref{GCAcentral}.  For Einstein gravity, $c_{\text{L}}=0, c_{\text{M}}=3/G$ \cite{Barnich:2006av}.

\subsection{Global Minkowski, null-orbifold and FSC}\label{GMNOFSC}
The zero mode solutions in \eqref{SolEOM}, describing some classical background of spacetime, are of particular interest.
With the standard  parameterization of the $S^1$ \be\label{fbtzid} \phi \sim \phi+2\pi\, , \ee
general classical solutions of Einstein gravity without cosmology constant takes the following form
\cite{Barnich:2010eb}
\bea\label{ClassSol}
ds^2&=&8G M du^2-2 dudr+8GJ du d\phi+r^2 d\phi^2 
\nonumber \\
&\rightarrow  & M du^2-2 dudr+ J du d\phi+r^2 d\phi^2.
 \label{fbtz}
\eea
where in the second line we have used the convention $8G=1$ which will be adopted throughout the paper. We will spell it out whenever it is necessary to restore $G$.  

Via holography, the identification (\ref{fbtzid}) specifies a canonical spatial circle where the BMSFT is defined on.
These classical flat-space backgrounds (\ref{ClassSol})  can be classified into three types:
\begin{itemize}
\item  $M=-1,J=0$: Global Minkowski.  The solution (\ref{fbtz}) (\ref{fbtzid}) cover the full three dimensional Minkowski. The holographic dual is the BMSFT defined on the cylinder.

\item   $M=J=0$: Null-orbifold. It was first constructed in string theory \cite{Horowitz:1990ap}. They are supposed to play the role of zero temperature BTZ,  being the holographic dual of  BMSFT on a torus with  zero-temperature and a fixed spatial circle.
\item  $M>0 $: Flat Space Cosmological solution (FSC). This was previously   studied in string theory as the shifted-boost orbifold of Minkowski spacetimes  \cite{Cornalba:2002fi,Cornalba:2003kd}. Their boundary dual is the thermal BMSFT at finite temperature.
\end{itemize}

The flat-space metric (\ref{ClassSol}) can also be written in the ADM form\footnote{The coordinates are related by
\be\label{ADMtoBMS}
t=u-\frac{r}{M}+\frac{r_c}{2M}\log\frac{r-r_c}{r+r_c},\qquad \varphi=\phi- \frac{r_c}{2\sqrt{M}}\log\frac{r-r_c}{r+r_c},\qquad r_c=\frac{J}{2\sqrt{M}}.
\ee}
\be
ds^2=-(\frac{J^2}{4r^2}-M) dt^2+\frac{dr^2}{(\frac{J^2}{4r^2}-M)}+r^2 (d\varphi +\frac{J}{2r^2} dt)^2\,.
\ee
This indicates that the flat-space \eqref{fbtz}  admit  a Cauchy horizon \cite{Barnich:2012aw} at
\be
r_H=|r_c|, \qquad r_c\equiv \frac{J}{2\sqrt{M}}\,.
\ee
The thermal circle of \eqref{fbtz} is given by
\begin{align}\label{thermalcircle}
\text{thermal circle}:~~(u,\phi)\sim (u+i\beta_u,\phi-i\beta_\phi)\,,
\end{align}
with
\begin{align}\label{betas}
\beta_u=\frac{\pi J}{M^{3/2}},\quad \beta_\phi= \frac{2\pi}{\sqrt{M}}.
\end{align}

The thermal entropy of the  horizon is given by the Bekenstein-Hawking  formula. Meanwhile, this thermal entropy also can be obtained from the dual field theory by considering a Cary-like counting of states. The equality of thermal entropy between bulk gravity and boundary field was shown in \cite{Bagchi:2012xr,Barnich:2012xq} 
\begin{align} 
S_{\text{FSC}}=S_{\text{BMSFT}}=\frac{2\pi r_c }{4G}\,.
\end{align}

 The flat-space metric (\ref{ClassSol}) with $M>0$  can also be brought into the Cartesian coordinate locally\footnote{We hope that the same symbol ``$t$''  with different meanings in \eqref{ADMtoBMS} and \eqref{BMStoCar} will not cause any confusions.}
\besub\label{BMStoCar}
\beqn
r &=& \sqrt{M (t^2- x^2)+r_c^2 } \,, \\
\phi &= & -\frac{1}{ \sqrt{M}} \log\frac{\sqrt{ M}(t-x)}{r+r_c} \,,\\
 u &=& \frac{1}{M}\Big( r - \sqrt{M} y -\sqrt{M}r_c \phi \Big)\,,
\eeqn
\eesub
satisfying $ds^2=-dt^2+dx^2+dy^2 $.

\subsection{Poincar\'{e} coordinates}
If we decompactify the angular direction $\phi$, the boundary theory will be put on the plane instead of cylinder. In particular, the resulting spacetime with $M=J=0$ is more like a flat version of AdS in Poincar\'{e} patch.
The coordinate transformation from the Poincar\'{e}  coordinate and the Cartesian coordinate is
\beqn
t &= &{l_\phi\over4}r +{2\over l_\phi}(u+\frac{r\phi^2}{2})  \,,\\
x &=&  ({l_u\over l_\phi}+r\phi)  \,,\\
y &= &{l_\phi\over4} r-{2\over l_\phi}(u+\frac{r\phi^2}{2})\,.
\eeqn
It is easy to check that
\bea
ds^2&=& -2 dudr +r^2 d\phi^2\,,\label{poincPoincar\'{e}are}\\
&=& -dt^2+dx^2+dy^2\,.
\eea the patch with $r\ge0,\, u\in (-\infty, \infty),\, \phi\in (-\infty,\infty)$
 covers the region $t+y\ge 0.$

The inverse transformation from Cartesian coordinate to the Poincar\'{e} to  is
 \bea
 r={2(t+y)\over l_\phi},\quad u={-(l_u-l_\phi x)^2+l_\phi^2 (t^2-y^2)\over 4 l_\phi (t+y)},\quad \phi={-l_u+l_\phi x\over2(t+y)}
\,. \eea
 As $r\rightarrow 0,$ both $u$ and $\phi $ will diverge except when $x\rightarrow {l_u\over l_\phi}$.

\subsection{Solutions with general spatial circle }
More generally, we can consider solutions locally with the same metric as (\ref{fbtz}), but with a spatial circle different from (\ref{fbtzid}).  More precisely,
\bea
ds^2&=&\tM d\tu^2-2 d\tu d\tdr+\tJ d\tu d\tphi+r^2 d\tphi^2\,, \label{tbTZ}\\
(\tu,\tphi) &\sim &(\tu+\Delta {\tilde u}, \tphi+\Delta {\tilde \phi})\,.  \label{tbTZid}
\eea
When we study the bulk extension of the Rindler transformations in section~\ref{GravityHEE}, we will encounter the bulk extensions of  $\tilde{\mathcal{B}}$, which are usually this kind of spacetime. Hereafter we will refer to (\ref{tbTZ}) as $\widetilde{\text{FSC}}$.
The proper length on Cauchy horizon is $ds^2=\left(\sqrt{\tilde{M}}d\tilde{u}+\tilde{r}_c d\tilde{\phi}\right)^2$. Integration along this proper length will give the length of the horizon, and the Bekenstein-Hawking entropy
\be\label{SBH}
S_{\text{BH}}=\frac{\ell_{\text{horizon}}}{4G}
=\frac{\sqrt{\tilde{M}}\Delta {\tilde u}+r_c \Delta {\tilde \phi}}{4G}\,,
\ee
where $\Delta {\tilde u}$ and $\Delta {\tilde \phi}$ represent the extension of the spacetime along the $\tilde u$ and $\tilde \phi$ direction.

\section{Entanglement entropy in field theory side} \label{FieldEE}
In this section we apply the generalized Rindler method to a general BMS field theory with arbitrary $c_{\text{L}}$ and $c_{\text{M}}$. As we will show below, our results in this section  agree  with the previous calculation using twist operators \cite{Bagchi:2014iea}. 
\subsection{Rindler transformations and the modular flow in BMSFT }\label{rindlerTranf}
In this section, following the guidelines we give in section~\ref{Rindlertransformations}, we derive the most general Rindler transformations in BMSFT.
According to property 1 in subsection \ref{Rindlertransformations}, the Rindler transformations should be a symmetry of the field theory, thus for BMSFT  it should be the following BMS transformations \cite{Barnich:2012xq}
\bea
\tilde{\phi}&=&f(\phi)\,,\\
\tilde{u}&=&\p f(\phi) u+g(\phi)\,,
\eea
where we use $\p$ to denote derivative with respect to $\phi$. Furthermore, this indicates the theory after Rindler transformation is also a BMS invariant field theory, which we call $\widetilde{\text{BMSFT}}$.
The inverse transformation can be written as
\bea
{\phi}&=&\tilde{f}(\tilde{\phi})\,,\\
{u}&=&\tp\tilde{f}(\tilde{\phi})\tilde{u}+\tilde{g}(\tilde{\phi})\,,
\eea
where \be\label{fgtilde} \tp\tilde{f}\p f =1,\quad \tilde{g}=-g\tp\tilde{f}\,,\ee
and $\tp$ denotes derivative with respect to $\tilde{\phi}$.
The property 4 of the Rindler transformation indicates that the vectors $\p_{\tilde{\phi}}$ and $\p_{\tilde{u}}$ have to be linear combinations of the global BMS$_3$ generators.
The conditions for $\p_{\tilde{\phi}}$ is
\bea
\p_{\tilde{\phi}}&=&\tp\tilde{f} \,\p_\phi+\tp^2\tilde{f}\tilde{u}\,\p_u+\tp\tilde{g} \,\p_u \\
&=&\tp\tilde{f} \,\p_\phi+\p_\phi\left(\tp\tilde{f}\right)u\,\p_u+\left(-\p_\phi\left(\tp\tilde{f}\right)\tilde{g}+\tp\tilde{g} \right)\p_u \nonumber\\
&=&\sum_{n=-1}^1( b_{n} L_n+ d_n M_n)\,. \label{dtphi}
\eea
which implies that the condition for $\p_{\tilde{u}}$ is automatically satisfied, as
\be\p_{\tilde{u}}=\tp\tilde{f}(\tilde{\phi}) \p_u=-\sum_{n=-1}^1 b_{n} M_n\,.\label{dtu}\ee

Note the BMS generators have the following general form (see \cite{Barnich:2012xq} or consider the  $u,\phi$ components of \eqref{BlkKilling} in the limit of $r\rightarrow \infty$)
\be\label{FieldKillingForm}
\sum_{n=-1}^1( b_{n} L_n+ d_n M_n) =(u\partial_\phi Y+T)\partial_u+Y\partial_\phi  \,,
\ee
 then we can get two differential equations
\bea
\tp\tilde{f}(\tilde{\phi}) &=&  Y(\phi)\,,\\
-\p_\phi\left(\tp\tilde{f}\right)\tilde{g}+\tp\tilde{g}&=& T(\phi)\,.
\eea
It is interesting to note
\be
\partial_\phi  \Big( \frac{\tilde g}{\tilde \partial \tilde f} \Big)
=\frac{-\p_\phi\left(\tp\tilde{f}\right)\tilde{g}+\partial_\phi \tilde g \cdot \tilde \partial \tilde f}{({\tilde \partial \tilde f})^2} 
=\frac{-\p_\phi\left(\tp\tilde{f}\right)\tilde{g}+ \tp\tilde{g}}{({\tilde \partial \tilde f})^2} 
=\frac{T}{Y^2}\,.
\ee
Furthermore by noting the relation in \eqref{fgtilde}, the previous differential equations can be simplified as 
\beqn
\partial_\phi f&=&\frac{1}{Y}\,,    \label{Eqf} \\
\partial_\phi g&=&- \frac{T}{Y^2}\,. \label{Eqg} 
\eeqn 

\subsubsection*{Rindler transformation on the plane}
Now we consider the BMSFT on the plane with zero temperature. In this case the symmetry generators of BMS$_3$ are
\bea
L_n&=& -u (n+1)\phi^n\p_u-\phi^{n+1}\p_\phi\,,\\
M_n&=&\phi^{n+1}\p_u\,,
\eea
where the $n=-1,0,1$ part form a subalgebra, and generate the global symmetries. By matching the general form of Killing vectors in \eqref{FieldKillingForm}, one can easily see that 
\be\label{YphiFcn}
Y=-b_{-1}-b_0\phi-b_1\phi^2 ,\qquad T= d_{-1 }+d_0 \phi +d_1 \phi^2\,.
\ee
Substituting them into \eqref{Eqf}, one gets 
\be
f=-\frac{2}{\sqrt{-b_0^2+4 b_1 b_{-1}}} \arctan \Big( \frac{b_0+2b_1 \phi}{\sqrt{-b_0^2+4 b_1 b_{-1}}}   \Big)+c_1\,.
\ee
Note that $c_1$ and $b_0$ shift the origin of $\phi$ or $\tilde{\phi}$, and therefore we can set $c_1=b_0=0$ without losing generality.  
Taking  
\be\label{b1b-1no}
 b_1={2\pi\over \tilde\beta_\phi l_\phi},\qquad \,b_{-1}=-{\pi l_\phi\over2\tilde\beta_\phi}\,,\ee
we get \be f(\phi)=\frac{\tilde\beta_\phi}{\pi }  \arctanh\left(\frac{2 \phi }{l_\phi }\right)\,.\ee
Plugging $\eqref{YphiFcn}$ and $b_{\pm 1,0}$ into  \eqref{Eqg} , we get
\be
g=\frac{-4d_{-1}+d_1 l_\phi^2}{4\pi^2 l_\phi } \tilde\beta_\phi^2\arctanh\left(\frac{2 \phi }{l_\phi }\right)
-\frac{\tilde\beta_\phi^2(4d_{-1}\phi+l_\phi^2 (d_0+d_1\phi^2))}{2\pi^2(l_\phi^2-4\phi^2)}
+c_2\,.
\ee
Taking 
\be\label{d1d-1no}
d_1=  \frac{2\pi l_u }{\tilde\beta_\phi l_\phi^2}-\frac{2\pi \tilde\beta_u}{l_\phi \tilde\beta_\phi^2 } ,
 \qquad d_{-1}= \frac{ \pi l_u }{2\tilde\beta_\phi  }    +  \frac{ \pi l_\phi \tilde\beta_u}{2 \tilde\beta_\phi^2 } ,
\ee
then, 
\be 
\tilde{u} ={2\tilde\beta_\phi(u l_\phi-l_u\phi )\over\pi( l_\phi^2-4\phi^2)}-{d_0\tilde\beta_\phi^2l_\phi^2\over 2\pi^2(l_\phi^2-4\phi^2)}-
\frac{\tilde\beta_u}{\tilde\beta_\phi}\tilde{\phi}+c_2\,.
\ee
The two parameters  $d_0$ and $c_2$ can be absorbed by a shift of $\tilde{u}$ and $u$. Finally, up to some trivial shifts, we get the most general coordinate transformations  
\besub\label{rindlertNO}
\bea
&&\tilde{\phi}=\frac{{\tilde{\beta}_\phi}}{\pi }  \tanh ^{-1}\left(\frac{2 \phi }{l_\phi }\right)\,,\\
&&\tilde{u}+{{\tilde\beta}_u\over {\tilde \beta}_\phi}\tilde{\phi}={2{\tilde{\beta}_\phi}(u l_\phi-l_u\phi )\over\pi( l_\phi^2-4\phi^2)}\,.\eea
\eesub
The above Rindler transformation satisfies the property 2 as it induces a thermal circle
\begin{align}
\text{thermal circle}:~~(\tilde{u},\tilde{\phi})\sim (\tilde{u}+i\tilde{\beta}_u,\tilde{\phi}-i\tilde{\beta}_\phi)\,.
\end{align}
Note that the subregion $\mathcal{D}$ of $\mathcal{B}$ which maps to $\tilde{\mathcal{B}}$ is a strip bounded by $\phi= - l_\phi/2$ and  $\phi= l_\phi/2$.

Rewritten in the original coordinate system, we get
\bea
\p_{\tilde \phi}&=& {\pi\over 2{\tilde{\beta}_\phi} l_\phi}   \left(( l_\phi ^2-4 \phi^2) \p_\phi \,+   ( l_ul_\phi +4 {l_u\over l_\phi }\phi^2-8 u \phi +{{\tilde\beta}_u\over {\tilde \beta}_\phi}  l_\phi ^2-4 {{\tilde\beta}_u\over {\tilde \beta}_\phi}  \phi^2) \p_u\right)\,,
 \\
\p_{\tilde{u}}&=& {\pi \over 2 {\tilde{\beta}_\phi}  l_\phi }  \left(l_\phi ^2-4 \phi^2\right)\p_u \,.\eea
Thus, the generator of modular flow is
\bea\label{ktLM}
k_t&=&-{\tilde{\beta}_\phi} (\p_{\tilde \phi} -{{\tilde\beta}_u\over {\tilde \beta}_\phi} \p_{\tilde{u}})
=-\sum_{n=-1}^1{\tilde{\beta}_\phi}(b_n L_n+(d_n+{{\tilde\beta}_u\over {\tilde \beta}_\phi} b_n) M_n)\,,
\\
&=&-{\pi\over 2  l_\phi}   \left(( l_\phi ^2-4 \phi^2) \p_\phi \,+   ( l_ul_\phi +4 {l_u\over l_\phi }\phi^2-8 u \phi ) \p_u\right)\,.
 \eea
It is easy to verify that at the points $(   \pm l_u/2 ,\pm l_\phi/2)$, $k_t=0$. Following our prescription \eqref{fixed}, this implies that $(  \pm l_u/2 , \pm l_\phi/2 )$ is that endpoints of the interval. Thus, we can naturally interpret $l_u, l_\phi $ as the extension of the interval along   $u$ and $\phi$ direction. 
 
The global BMS$_3$ generators  $L_n, M_n$ have bulk extensions, which are just the Killing vectors \eqref{NOKilling} of the flat-space \eqref{fbtz} in Poincar\'{e} patch. Substitute \eqref{NOKilling} into \eqref{dtphi} and \eqref{dtu}, the modular generator \eqref{ktLM} can also be extended to the bulk and is given by
\begin{align}
k_t^{\text{bulk}}
=&- {\pi\over 2  l_\phi}   \left(( l_\phi ^2-4 \phi^2+\frac{8(l_\phi u-l_u\phi)}{l_\phi r}) \p_\phi \,+   ( l_ul_\phi +4 {l_u\over l_\phi }\phi^2-8 u \phi ) \p_u
+  (\frac{8l_u}{l_\phi}  +8 r\phi)\partial_r      \right)  \,.
\end{align}

Following the similar analysis, we give the construction of Rindler transformations in BMSFT with a thermal or spatial circle in Appendix~\ref{RindlerTsFSC}.  We also calculate the modular flow and its bulk extension in these two cases.

\subsection{Entanglement entropy for BMSFT}
In this subsection we consider zero temperature BMSFT on the plane, finite temperature BMSFT, and BMSFT on a cylinder respectively, and calculate the entanglement entropy of the following interval in these BMSFTs, 
\be\label{interval}
\mathcal{A}:\qquad (  - l_u/2 ,  - l_\phi/2  ) \rightarrow (  l_u/2 , l_\phi/2 )\,,
\ee
 where the arrow means a line connecting the two endpoints.  After the Rindler transformation, the entanglement entropy equals to the thermal entropy of the $\widetilde{\text{BMSFT}}$ on $\tilde{\mathcal{B}}$, which can be calculated via a Cardy-like formula. Since the extension of $\tilde{\mathcal{B}}$ is essentially infinite, we need to introduce the cutoffs to regulate the interval
\begin{align}
\mathcal{A}_{\text{reg}}:\qquad (  - l_u/2+\epsilon_u ,  - l_\phi/2+\epsilon_\phi  ) \rightarrow (  l_u/2-\epsilon_u , l_\phi/2-\epsilon_\phi )\,.
\end{align}
 In \cite{Bagchi:2012xr,Barnich:2012xq}, the Cardy-like formula for BMSFT is derived from the modular invariance of the theory, which is inherited from the modular invariance of CFT$_2$ under flat limit. In appendix~\ref{CardyThermal} we re-derive the Cardy-like formula using the BMS symmetry only. 
 
The manifold $\tilde{\mathcal{B}}$ can be considered as a torus with the following identifications
\be\label{torusab}
(\tilde u,\tilde \phi ) \sim (\tilde u+i \bar{a},\tilde \phi-i a )
\sim (\tilde u+2\pi \bar{b},\tilde \phi-2\pi b)\,,
\ee
where $(a,\bar{a})$ parametrize a thermal circle and $(b,\bar{b})$ parametrize a spatial circle. We find that (see appendix~\ref{CardyThermal}), under some regime \eqref{cardyregion}, the thermal entropy can be calculated by 
\bea\label{SEE}
S_ {\bar{b}|b}({\bar{a}|a} )
&=& - \frac{\pi^2}{3}  \Big(c_{\text{L}} \frac{ b  }{a}
+ c_{\text{M}}\frac{  (\bar a b-a \bar b)  }{a^2}  \Big)\,.
\eea

\subsubsection{Zero temperature BMSFT on the plane}
After performing the Rindler transformation \eqref{rindlertNO}, the image of this regularized interval $\mathcal{A}_{\text{reg}}$ is
\bea
\mathcal{I_{\text{reg}}}:
&&\quad  -( {\Delta \tilde u\over2}  ,  {\Delta \tilde \phi\over2}    )   \rightarrow ( {\Delta \tilde u\over2} , {\Delta \tilde \phi\over2} )\,,\\
&&\Delta \tilde \phi=  \frac{{\tilde{\beta}_\phi}}{\pi}\log \frac{l_\phi}{\epsilon_\phi},\quad \Delta \tilde u= \frac{{\tilde{\beta}_\phi} l_u }{\pi l_\phi}-\frac{{\tilde{\beta}_\phi} \epsilon_u}{\pi \epsilon_\phi}-\frac{{\tilde{\beta}_u}}{\pi}\log \frac{l_\phi}{\epsilon_\phi}\,, \nonumber
  \eea
where we have neglected irrelevant terms $\mathcal{O}(\epsilon_u,\epsilon_\phi)$ and terms $\mathcal{O}(\epsilon_u^2)/\epsilon_\phi$, but keep  terms of order ${\epsilon_u \over \epsilon_\phi}$. 
The endpoint effects are expected to be negligible for this large interval, thus we can identify the endpoints to form a spatial circle. The same story happens in the later two cases. Together with the thermal circle induced by the Rindler transformation \eqref{rindlertNO}, $\tilde{\mathcal{B}}$ can be considered as a torus parametrized by
\beqn
 a= {\tilde{\beta}_\phi}&,\qquad&   \bar{a}= {\tilde{\beta}_u}\,, \\
 2\pi b=- \Delta \tilde \phi &,\qquad&
  2\pi \bar{b}=\Delta \tilde u\,.
\eeqn
The canonical torus parameters  defined by \eqref{ModularParameter} are
\be
\hat\beta_\phi=- \frac{2 \pi^2}{\log \frac{l_\phi}{\epsilon_\phi}} , \qquad 
\hat\beta_u=-\frac{\hat\beta_\phi^2 }{2\pi^2}  \Big( \frac{l_u}{l_\phi}  - \frac{  \epsilon_u}{\epsilon_\phi}    \Big)\,.
\ee
thus satisfy the regime \eqref{cardyregion} for \eqref{SEE}.

Substituting these quantities into the entropy formula \eqref{SEE} derived before, we can obtain the entropy
\be\label{Seeplane}
S_{\text{EE}}=  \frac{c_{\text{L}}}{6}  \log \frac{l_\phi}{\epsilon_\phi}
+ \frac{c_{\text{M}}}{6}   \Big( \frac{l_u}{l_\phi} -
\frac{\epsilon_u}{\epsilon_\phi} \Big)\,.
\ee
This agrees with the result in literature \cite{Bagchi:2014iea} when we set $\epsilon_u=0$. In general, we would like to keep the cutoff related terms. It is interesting to note that $c_{\text{L}}$ term is the same as that in CFT. This is not surprising since the $\mathcal L_n$ generators satisfy the chiral part of the Virasoro algebra.

\subsubsection{Finite temperature BMSFT}
The temperature of the quantum field theory is dictated by the periodicity along the imaginary axis of time, namely $u\sim u+i\beta $. More generally, we can have the following thermal identification
\be
(  u,\phi) \sim ( u+i\beta_u,\phi-i\beta_\phi)\,.
\ee

The Rindler transformation for  thermal BMSFT is given in  \eqref{rindlertFSC}, which in particular respects the above thermal circle. Following the similar steps, we find the torus $\tilde{\mathcal{B}}$ is parametrized by
\beqn
 a=  {\tilde{\beta}_\phi}&,\qquad& \bar{a}=  {\tilde{\beta}_u} \,,\\
 2\pi b=- \frac{{\tilde{\beta}_\phi}}{\pi}\zeta &,\qquad&
  2\pi \bar{b}=  -\frac{{\tilde{\beta}_u}}{ \pi }\zeta +
  \frac{{\tilde{\beta}_\phi}}{ \pi \beta_\phi} \Big[ \pi \Big(l_u+\frac{\beta_u}{\beta_\phi}l_\phi \Big) \coth\frac{\pi l_\phi}{\beta_\phi}
  -\beta_u\Big]\,,
\eeqn
where  $\zeta=\log \Big( \frac{\beta_\phi}{\pi   \epsilon_\phi}\sinh\frac{\pi l_\phi}{\beta_\phi} \Big)$. 

One can check that the canonical torus parameters \eqref{ModularParameter} also satisfy the regime \eqref{cardyregion}, thus the EE can be obtained from \eqref{SEE}
\be\label{Seethermal}
S_{\text{EE}}= \frac{c_{\text{L}}}{6}  \log\Big( \frac{\beta_\phi}{\pi   \epsilon_\phi}\sinh\frac{\pi l_\phi}{\beta_\phi} \Big)
+\frac{c_{\text{M}}}{6}   \frac{1}{  \beta_\phi} \Big[ \pi \Big(l_u+\frac{\beta_u}{\beta_\phi}l_\phi \Big)
\coth  \Big(  \frac{\pi l_\phi}{\beta_\phi}  \Big)
  -\beta_u\Big]
-  \frac{c_{\text{M}}}{6}
\frac{\epsilon_u}{\epsilon_\phi}\,,
\ee
coinciding with the result in \cite{Basu:2015evh}.

\subsubsection{Zero temperature BMSFT on the cylinder}
Consider the cylinder with periodicity $\phi\sim \phi+2\pi$, we should apply the Rindler transformation \eqref{rindlertGM}. Similarly we find $  \mathcal{\tilde B}$  can be considered as a torus parametrized by  
\beqn
  a=  {\tilde{\beta}_\phi}&,\qquad&  \bar{a}=   {\tilde{\beta}_u} \,,\\
 2\pi b=- \frac{{\tilde{\beta}_\phi}}{\pi}\zeta &,\qquad&
  2\pi \bar{b}=-\frac{{\tilde{\beta}_u}}{ \pi}\zeta +\frac{{\tilde{\beta}_\phi} l_u \cot(l_\phi/2)}{2 \pi  }-\frac{{\tilde{\beta}_\phi} \epsilon_u}{ \pi \epsilon_\phi}\,,
\eeqn
where $\zeta=\log \Big( \frac{2}{\epsilon_\phi}\sin\frac{l_\phi}{2} \Big)$. When $l_\phi<\pi$, again the canonical parameters defined in \eqref{ModularParameter} satisfies the regime \eqref{cardyregion} for \eqref{SEE}. Thus the entropy can be calculated by \eqref{SEE},
\be\label{Seecylinder}
S_{\text{EE}} =  \frac{c_{\text{L}}}{6}    \log \Big( \frac{2}{\epsilon_\phi}\sin\frac{l_\phi}{2} \Big)
+ \frac{c_{\text{M}}}{12} \Big(   l_u \cot\Big(\frac{l_\phi}{2}\Big)   - \frac{2\epsilon_u}{\epsilon_\phi}     \Big)\,.
\ee

When the interval is very small, i.e. $ l_\phi  \rightarrow 0 $, $
S^{cyl}_{\text{EE}} \rightarrow S^{plane}_{\text{EE}}
 $, which is just the EE on the plane. This is reasonable, since when the interval is very small, whether the space is compact or not is expected to be irrelevant.

\section{Holographic entanglement entropy in gravity side}\label{GravityHEE}
\subsection{The strategy }\label{strategy}
In this section we extend the field side story to the bulk. On the field theory side, the Rindler transformations map the entanglement entropy for BMSFT to the thermal entropy of $\widetilde{\text{BMSFT}}$ . According to  section~\ref{BMSGCA}, this story should have a bulk description. We start from BMSFT and its gravity dual, the flat-space \eqref{fbtz}. Then our task is to find a transformation from \eqref{fbtz} to a new spacetime, which is usually a $\widetilde{\text{FSC}}$ \eqref{tbTZ} in tilde coordinates, such that its dual boundary field theory is just the $\widetilde{\text{BMSFT}}$  in the field theory side story. According to the flat holography, the Bekenstein-Hawking entropy of $\widetilde{\text{FSC}}$  equals to the thermal entropy of $\widetilde{\text{BMSFT}}$, thus, reproduces the entanglement entropy for BMSFT.

To make sure that the gravity side story reproduces the field theory side story on the boundary, we expect $\widetilde{\text{FSC}}$  to satisfy the following requirements.
\begin{itemize}
\item
The bulk transformations from flat-space \eqref{fbtz} to $\widetilde{\text{FSC}}$ should be a bulk extension of the Rindler transformation, thus reproduce the field side story on the boundary.

 \item
 The asymptotic structures of $\widetilde{\text{FSC}}$  should satisfy the BMS$_3$ boundary conditions locally, and have the same thermal and spatial periodicities as the boundary field theory $\widetilde{\text{BMSFT}}$ .
\end{itemize}
More explicitly we expect the metric of $\widetilde{\text{FSC}}$  to be in the form of (\ref{fbtz}).  The second requirement is necessary, since we expect $\widetilde{\text{FSC}}$  to be the gravity dual of $\widetilde{\text{BMSFT}}$ .

Based on our discussions on the Rindler transformations in the field theory side story, we find that the vectors $\p_{\tilde{u}}$ and $\p_{\tilde{\phi}}$ should be the following linear combinations of the global BMSFT generators
\begin{align}\label{tphitu}
\p_{\tilde{\phi}}=\sum_{n=-1}^1( b_{n} L_n+ d_n M_n) \,,
\qquad
\p_{\tilde{u}}=-\sum_{n=-1}^1 b_{n} M_n\,.
\end{align}
This relation can be naturally extended to the bulk by simply replacing the global generators in BMSFT with their bulk counterparts, i.e. the Killing vectors in flat-space. These Killing vectors are explicitly given by (\ref{NOKilling}) for Poincar\'{e}, and by (\ref{KillingFSC}) for FSC and global Minkowski. The first requirement indicates that, in order to reproduce the Rindler transformation on the boundary, we should choose the same coefficients $b_{0,\pm 1}$ and $d_{0,\pm 1}$ as in the field theory side. The second requirement gives the Bondi gauge conditions for the new coordinates
\begin{align}\label{condition1}
g_{\tilde{u}\tilde{u}}= \p_{\tilde{u}}\cdot \p_{\tilde{u}}\equiv\tilde{M}\,,\qquad g_{\tilde{u}\tilde{\phi}}=\p_{\tilde{u}}\cdot\p_{\tilde{\phi}}\equiv\tilde{J}\,,\qquad g_{\tilde{\phi}\tilde{\phi}}=\p_{\tilde{\phi}}\cdot\p_{\tilde{\phi}}\equiv \tilde{r}^2\,,
 \\\label{condition2}
 g_{\tilde{r}\tilde{r}}=\p_{\tilde{r}}\cdot\p_{\tilde{r}}=0\,,\qquad g_{\tilde{r}\tilde{\phi}}=\p_{\tilde{r}}\cdot\p_{\tilde{\phi}}=0\,,\qquad g_{\tilde{u}\tilde{r}}=\p_{\tilde{u}}\cdot\p_{\tilde{r}}= g_{\tilde{u}\tilde{r}}(\tilde{r})\,,
 \end{align}
where the inner products are calculated with the old metric of flat-space. The constants $\tilde{M}$ and $\tilde{J}$ are determined by the coefficients in \eqref{tphitu} and can be regarded as the mass and angular momentum of the new spacetime $\widetilde{\text{FSC}}$ . Note that, in the bulk $\p_{\tilde{u}}$ and $\p_{\tilde{\phi}}$ are two commuting Killing vectors, the new metric only depend on the third metric $\tilde{r}$. So the Bondi gauge conditions \eqref{condition1} \eqref{condition2} are consistent with \eqref{tphitu}. We define the new radial coordinate $\tilde{r}$ with the third equation in (\ref{condition1}), and the third equation in (\ref{condition2}) requires $g_{\tilde{u}\tilde{r}}$ only depends on $\tilde{r}$. Combing these vectors $(\p_{\tilde u}, \p_{\tilde \phi}, \p_{\tilde r})$ together, we get a matrix, which is essentially the Jacobian matrix between the old and new coordinate systems. Solving all these conditions will give the bulk coordinate transformation, as well as the unknown metric component $g_{\tilde{u}\tilde{r}}(\tilde{r})$. As expected, we always get $g_{\tilde{u}\tilde{r}}(\tilde{r})=-1$, so the new metric is in the form of \eqref{fbtz}.

The bulk transformation can be regarded as a quotient on flat-space without doing identification for the new coordinates. Similar strategy has been successfully applied on warped AdS$_3$ and AdS$_3$ spacetimes with certain boundary conditions in \cite{Song:2016gtd} to calculate holographic entanglement entropy, and both the results fulfill their field theory side expectations.

We will use the method elaborated above to perform quotient for  Poincar\'{e} coordinate in subsection \ref{NOHee}. Then a short-cut method of quotient for FSC and global Minkowski in Einstein gravity will be used for calculating HEE in section~\ref{FSCHee} and  section~\ref{MinkHee}, respectively. Note that, in Einstein gravity, the central charges of the dual BMSFT are given by $c_{\text{L}}=0,~c_{\text{M}}=3/G$, with $c_{\text{L}}$ vanished. To investigate the holographic entanglement entropy contributed from the $c_{\text{L}}$ term, we will consider the topologically massive gravity (TMG) in section~\ref{TMG}.

\subsection{Poincar\'{e} coordinate}\label{NOHee}
In Poincar\'{e} coordinates, the coefficients in (\ref{tphitu}) are given by \eqref{b1b-1no} and \eqref{d1d-1no}, while the bulk Killing vectors are given by (\ref{NOKilling}). We also need to use the following relations
\begin{align}\label{betaMJ}
{\tilde{\beta}_\phi}= \frac{2\pi}{\sqrt{\tilde M}}, \qquad {{\tilde\beta}_u\over {\tilde \beta}_\phi}=\rzero\,,
\end{align}
which is just \eqref{betas} in $\widetilde{\text{FSC}}$.   Following the steps outlined in section~\ref{strategy}, we get the coordinate transformation between the ``Poincar\'{e}'' coordinates and $\widetilde{\text{FSC}}$,
\beqn
\tilde r &=& \sqrt{{{\tilde M}\over 16 l_\phi^2}\Big(8u-4l_u+r(l_\phi-2\phi )^2\Big) \Big(8u+4l_u+r(l_\phi+2\phi )^2\Big) +{{\tilde J}^2\over 4 \tilde M}} \,,\\
\tilde \phi &=& -\frac{1}{\sqrt{\tilde{M}}} \log {\sqrt{\tilde M}\over 4 l_\phi} \Big(\frac{8  u-4l_u+r(l_\phi-2\phi )^2} {{\tilde r}+{\tilde J /( 2\sqrt{\tilde M})} }\Big)\,,
 \\
\tilde u &=& {\tilde r \over \tilde M}+ {1\over 4l_\phi\sqrt{\tilde M}}\big( 8u+4 r\phi^2-r l_\phi^2\big)  -{\tilde J\over 2\tilde M} \tilde \phi\,.
\eeqn
It is easy to check that
\beqn
ds^2&=& \tilde M d\tilde u^2-2 d\tilde ud\tilde r +\tilde J d\tilde ud\tilde \phi+\tilde r^2 d\tilde\phi^2\,,
\\
&=&  -2 dudr +r^2 d\phi^2.
\eeqn
On the boundary, i.e. $r\rightarrow \infty$, the coordinate transformation becomes
\beqn\label{NullOrbBdy}
\tilde \phi &=&  \frac{2}{\sqrt{\tilde M}}\arctanh \frac{2\phi}{l_\phi}  \,,\\
\tilde  u &=&{4\over \sqrt{\tilde M}}  \frac{ (u l_\phi - l_u \phi)}{(l_\phi^2-4\phi^2)}
  -\frac{\tilde{r}_c }{\sqrt{\tilde{M}}}\tilde{\phi} \,,
\eeqn
which reproduces the Rindler transformation \eqref{rindlertNO} in the field theory side story after using the identification \eqref{betaMJ}.

By virtue of the flat holography, we can calculate the thermal entropy of $\widetilde{\text{BMSFT}}$  by the Bekenstein-Hawking entropy $S_{\text{BH}}$ of $\widetilde{\text{FSC}}$, 
\be
 S_{\text{BH}} =\frac{\sqrt{\tilde M}\Delta \tilde u+\tilde{J}/(2\sqrt{\tilde{M}}) \Delta \tilde \phi}{4G}~,
\ee
where $\Delta \tilde{u}$ and $\Delta\tilde{\phi}$ are the extension of the new coordinates. Since these two quantities are essentially infinity, we need to   introduce   two cutoffs $\epsilon_{u}$ and $\epsilon_{\phi}$, as was discussed on the field theory side before,  and consider the regularized interval 
\begin{align}
-\frac{l_{\phi}}{2}+\epsilon_{\phi}<\phi<\frac{l_{\phi}}{2}-\epsilon_{\phi}\,,\qquad -\frac{l_u}{2}+\epsilon_u<u<\frac{l_u}{2}-\epsilon_u\,.
\end{align}
Then it is easy to find that
\begin{align}
\Delta \tilde{u}=&\frac{2}{\sqrt{\tilde{M}}}\left(\frac{l_u}{l_{\phi}}-\frac{\epsilon_u}{\epsilon_{\phi}}\right)-\frac{\tilde{J}}{2\tilde{M}}\Delta\tilde{\phi}\,,
\\\label{noextension}
 \Delta \tilde{\phi}=&\frac{2}{\sqrt{\tilde{M}}}\log \frac{l_{\phi}}{\epsilon_\phi}\,.
\end{align}
Very straightforwardly, we    get the holographic entanglement entropy for BMSFT,
\begin{align}
S_{\text{\text{HEE}}}=S_{\text{BH}}=\frac{1}{2G}\left(\frac{l_u}{l_{\phi}}-\frac{\epsilon_u}{\epsilon_{\phi}}\right)\,.
\end{align}
As expected, this result agrees with the field theory side result \eqref{Seeplane}  after inserting the central charges of BMSFT dual to Einstein gravity $c_{\text{L}}=0,~c_{\text{M}}=\frac{3}{G}$.


\subsection{FSC }\label{FSCHee}
In principle, the strategy we applied on the Poincar\'{e} case can be   generalized to FSC and global Minkowski. However, the differential equations in these two cases are too complicated to solve. So we have to  find some short-cuts to simplify the calculation. Since the spacetimes we are studying are  locally flat, we can cast them into Cartesian coordinates. This yield many simplifications, since  in Cartesian coordinates the quotient from flat-space to $\widetilde{\text{FSC}}$  is essentially a Poincar\'{e} transformation. Then, by transforming back to the Bondi gauge, we get the quotient from flat-space to $\widetilde{\text{FSC}}$.

More specifically, we can associate the flat-space (FSC) with a Cartesian coordinate system $(x,y,t)$  through
\besub\label{tocartsian} 
\beqn
  r &=& \sqrt{M(t^2-x^2)+r_c^2 } \,,
  \\
  \phi &=-&\frac{1}{ \sqrt{  M}} \log\frac{\sqrt{M}(t-x)}{r+r_c}\,,
   \\
  u &=& \frac{1}{ M}\Big(   r - \sqrt{M} y-\sqrt{  M} r_c \phi \Big)\,.
\eeqn
\eesub
Similarly, with $M,~r_c$ replaced by $\tilde{M}, ~\tilde{r}_c$, we can associate the new spacetime $\widetilde{\text{FSC}}$ with a Cartesian coordinate system $(\tilde{x},\tilde{y},\tilde{t})$.
Our steps can be summarized by the following equations
 \begin{align}
ds^2=& M du^2-2 du dr+2 r_c\sqrt{M} du d\phi+r^2d\phi^2
\cr
 =&-dt^2+dx^2+dy^2 
 \cr
 =&-d\tilde t^2+d\tilde x^2+d\tilde y^2 
 \cr
 =&\tilde M d\tilde u^2-2 d\tilde u d\tilde r+2 \tilde r_c\sqrt{\tilde M} d\tilde u d\tilde \phi+\tilde r^2d\tilde \phi^2 \,.
 \end{align}
 Namely, we first consider the quotient between $(x,y,t)$ and $(\tilde{x},\tilde{y},\tilde{t})$ which is essentially a Poincar\'{e} transformation, then we can express their relations in  Bondi gauge with some coordinate substitutions. In order to reproduce the Rindler transformation (\ref{rindlertFSC}) asymptotically in Bondi gauge, we need to choose the Poincar\'{e} transformation properly. For our purpose, the relevant Poincar\'{e} transformation can be found as
\besub\label{QuotientCartesian}
\beqn
 \tilde t&=& t \cosh\eta  - y \sinh \eta\,,  \\
 \tilde y&=&y \cosh\eta  -  t \sinh \eta\,,  \\
 \tilde x&=& s_0+x\,,
\eeqn
\eesub
where the rapidity of boost along the $y$ direction, and the translation along the $x$ direction   are given by
\begin{align}\label{etas0}
\eta = \log \left(\coth \left(\frac{l_{\phi} \sqrt{M}}{4}\right)\right)\,,\qquad s_0=\frac{1}{2} (\sqrt{M}l_u +r_c l_\phi) \csch{(\sqrt{M}l_\phi/2)}\,.
\end{align}

Combining the coordinate transformations (\ref{tocartsian}) and  its tilde version as well as (\ref{QuotientCartesian}), it is straightforward to get the quotient from flat-space to $\widetilde{\text{FSC}}$ in Bondi gauge,  a coordinate transformation from $(u,r,\phi)$ to $(\tilde u, \tilde r, \tilde \phi)$\besub\label{RindlerFSC}
\begin{align} 
\tilde r=&\sqrt{R_+R_-+\tilde r_c^2}  \,,
\qquad \quad \\
\tilde \phi =& -\frac{1}{\sqrt{\tilde  M}}\log \Big( \frac{R_+}{\tilde r+\tilde r_c}  \Big)\,,
\qquad\label{phitilde}\\
\tilde u=&\frac{1}{  \tilde M } \Bigg[ \tilde r-\sqrt{\tilde M} \tilde r_c \tilde \phi 
   + \frac{\sqrt{\tilde M}}{ \sqrt{M} \sinh{(\sqrt{M}l_\phi/2)}}
\nonumber \\& \qquad\qquad \quad 
 \times   \Big( (-r+Mu+\sqrt{M} r_c\phi) \cosh\frac{\sqrt{M}l_\phi}{2}+(r-r_c) \cosh(\sqrt{M}\phi)
    \Big) \Bigg]\, ,
\label{utilde}\end{align}
\eesub
where
\beqn
R_\pm&=&\frac{\sqrt{\tilde M}}{ \sqrt{M} \sinh{(\sqrt{M}l_\phi/2)}} \Bigg[ M(u \mp \frac{l_u}{2}  )   +2 r\sinh^2\Big(\frac{\sqrt{M} ( 2\phi \mp l_\phi) }{4}\Big)
\nonumber \\&&\qquad\qquad\qquad\qquad\qquad
+r_c \sqrt{M}   (\phi \mp \frac{l_\phi}{2}) - r_c \sinh \Big(\frac{\sqrt{M} ( 2\phi \mp l_\phi) }{2} \Big)
 \Bigg]~.
\eeqn

The asymptotic behavior of the bulk transformation is  given by
\beqn\label{FSCbdy}
\tilde \phi &=&\frac{2}{\sqrt{\tilde M}}  \arctanh\left[\frac{\tanh (\sqrt{M}\phi/2)}{ \tanh (\sqrt{M}l_\phi/4) }\right]\,,  \\
\tilde u
&=&   \frac{\sqrt{M}u+r_c\phi-s_0\sinh(\sqrt{M}\phi)}
{\sqrt{\tilde M}\Big(\coth(\sqrt{M}l_\phi/2)-\cosh(\sqrt{M}\phi)\csch(\sqrt{M}l_\phi/2) \Big)}
-\frac{\tilde r_c}{\sqrt{\tilde M}} \tilde \phi\,,
\eeqn
which exactly matches the Rindler transformation (\ref{rindlertFSC}) in the field theory side story if we use \eqref{betas} and \eqref{betaMJ}. Introducing two regulating parameters $\epsilon_u$ and $\epsilon_\phi$, it is easy to determine the extension $\Delta \tilde{u}$ and $\Delta \tilde{\phi}$, which are given by
\begin{align}
\Delta\tilde{u}=&\frac{(l_u M+l_{\phi}\sqrt{M}r_c)\coth\left(\frac{l_{\phi}\sqrt{M}}{2}\right)-2 r_c}{\sqrt{M\tilde{M}}}-\frac{\tilde{r}_c}{\sqrt{\tilde{M}}}\Delta\tilde{\phi}-\frac{2\epsilon_u}{\sqrt{\tilde{M}}\epsilon_{\phi}}\,,
\\\label{fscextension}
\Delta\tilde{\phi}=&\frac{2}{\sqrt{\tilde{M}}}\log\left(\frac{2\sinh\frac{l_{\phi}\sqrt{M}}{2}}{\sqrt{M}\epsilon_{\phi}}\right)\,.
\end{align}

The holographic entanglement entropy is given by the Bekenstein-Hawking entropy of $\widetilde{\text{FSC}}$,
\be \label{sbht}
S_{\text{BH}} =\frac{\sqrt{\tilde M}\Delta \tilde u+\tilde r_c \Delta \tilde \phi}{4G}
 =\frac{1}{4G} \Big[ \sqrt{M} \Big( l_u+\frac{J}{2M}l_\phi \Big) \coth\frac{\sqrt{M}l_\phi}{2}-\frac{J}{M}-\frac{2\epsilon_u}{\epsilon_\phi}\Big]\,,
 \ee
which, according to \eqref{betas}, gives that
\be
S_{\text{\text{HEE}}}= S_{\text{BH}}=\frac{1}{2G } \Big[ \frac{\pi}{\beta_\phi} \Big(l_u+\frac{\beta_u}{\beta_\phi}l_\phi \Big)
\coth  \Big(  \frac{\pi l_\phi}{\beta_\phi}  \Big)
  -\frac{\beta_u}{\beta_\phi}-\frac{\epsilon_u}{\epsilon_{\phi}}\Big]\,.
\ee
As expected, this agrees with field theory side result (\ref{Seethermal}) with $c_{\text{L}}=0,~c_{\text{M}}=\frac{3}{G}$.

\subsection{Global Minkowski} \label{MinkHee}
For global Minkowski, we have conventional Cartesian coordinate system and the $\phi$ coordinate is compactified as $\phi\sim\phi+2\pi$. We define the our Cartesian coordinates in the following way,
\begin{align}\label{tocartsian2}
t=r+u, \qquad x=r \cos\phi, \qquad y=r\sin\phi\,,\qquad  \phi\sim \phi+2\pi \,,
\end{align}
As in the FSC case, the quotient here can also be performed in the Cartesian coordinate systems,
\begin{align}\label{MinkMetric}
ds^2=&-du^2-2 du dr+ r^2d\phi^2=-dt^2+dx^2+dy^2
    \cr
    =&-d\tilde{t}^2+d\tilde{x}^2+d\tilde{y}^2\
    \cr
    =&\tilde M d\tilde u^2-2 d\tilde u d\tilde r+2 \tilde r_c\sqrt{\tilde M} d\tilde u d\tilde \phi+\tilde r^2d\tilde \phi^2 \,.
\end{align}
Note that,   $(\tilde{t},\tilde{x},\tilde{y})$  is related to $(\tilde{u},\tilde{\phi},\tilde{r})$ in the same way as \eqref{tocartsian}.
Again the relevant Poincar\'{e} transformation is given by \eqref{QuotientCartesian} with $\eta$ and $s_0$ given by by \eqref{etas0}($M=-1$ and $r_c=0$ in this case).  The final quotient is given by (\ref{RindlerFSC}) with $M=-1,\,r_c=0$.
Asymptotically, the quotient reproduces the Rindler transformation \eqref{rindlertGM}. This together with the two regulation parameters $\epsilon_u$ and $\epsilon_{\phi}$ determines the extension of the horizon
\begin{align}\label{tuextensionGM}
\Delta \tilde{u}=&\frac{l_u}{\sqrt{\tilde{M}}}\cot\frac{l_\phi}{2}-\frac{\tilde{r}_c}{\sqrt{\tilde{M}}}\Delta\tilde{\phi}-\frac{2\epsilon_u}{\sqrt{\tilde{M}}\epsilon_\phi}\,,
\\\label{gmextension}
 \Delta \tilde{\phi}=&\frac{2}{\sqrt{\tilde{M}}}\log\left(\frac{2\sin\frac{l_{\phi}}{2}}{\epsilon_{\phi}}\right)\,.
 \end{align}
For Einstein gravity, this directly gives the holographic entanglement entropy
\begin{align}\label{Scylinder}
S_{\text{\text{HEE}}}= S_{\text{BH}}=\frac{\sqrt{\tilde{M}}\Delta \tilde{u}+\tilde{r}_c \Delta \tilde{\phi}}{4G}= \frac{1}{4G}  l_u\left|\cot \frac{l_\phi}{2}-\frac{2\epsilon_u}{\epsilon_\phi}\right|\,,
\end{align}
which is exactly the same as the field theory result \eqref{Seecylinder} with $c_{\text{L}}=0,~c_{\text{M}}=\frac{3}{G}$.

\section{The geometric description for holographic entanglement entropy}\label{geometry}
In this section, we provide a geometric description of the holographic entanglement entropy in global Minkowski.  
As discussed in section 2, there are two aspects, one from the explicit Rindler transformation in the bulk, and the other from the fixed points of the modular generator $k_t$.

\subsection{Three special curves}
From the bulk Rindler transformation, a natural step is to find the surface in the global Minkowski which is mapped to the  Cauchy horizon in $\widetilde{\text{FSC}}$. To make this subsection more self-contained, we rewrite (\ref{RindlerFSC}) with $M=-1,\, r_c=0$ below,
\besub\label{MinkBulk}
\beqn
\tilde r&=&\sqrt{R_+R_-+\tilde r_c^2}  \,,
\qquad \quad \\
\tilde \phi &=& -\frac{1}{\sqrt{\tilde  M}}\log \Bigg[ {R_+ \over \sqrt{R_+R_-+ {\tilde r}_c^2}+\tilde r_c} \Bigg]\,,
\qquad\label{phitilde}\\
\tilde u&=&\frac{1}{  \tilde M } \Big[ \tilde r-\sqrt{\tilde M} \tilde r_c \tilde \phi 
   -  \sqrt{\tilde M} \Big(r\cos\phi\csc(\frac{l_\phi}{2})-(r+u)\cot(\frac{l_\phi}{2})   \Big) \Bigg]\,.
\label{utilde}
\eeqn
\eesub
where \bea R_\pm= {\sqrt{\tilde M}\over  \sin ({l_\phi\over2})}\Big(2r\sin^2{l_\phi\mp 2\phi\over4}+u\mp {l_u\over2}\Big) \eea
The condition $\tilde r=\tilde r_c$ implies $R_+R_-=0$, whose solutions define two surfaces, 
\bea
\mathcal{N}_+:&& R_+=0\\
\mathcal{N}_-:&& R_-=0 \eea
with normal vectors \be n_\pm=\nabla_\mu R_\pm={\sqrt{\tilde M}\over  \sin ({l_\phi\over2})} \Big( 1,\, -r\sin({l_\phi\over2}\mp\phi),\,1-\cos({l_\phi\over2} \mp\phi)\Big)\ee
It is easy to verify that $n_+^2=n_-^2=0$, and hence both surfaces are null. 

\begin{figure}[h] 
   \centering
    \includegraphics[width=0.85\textwidth]{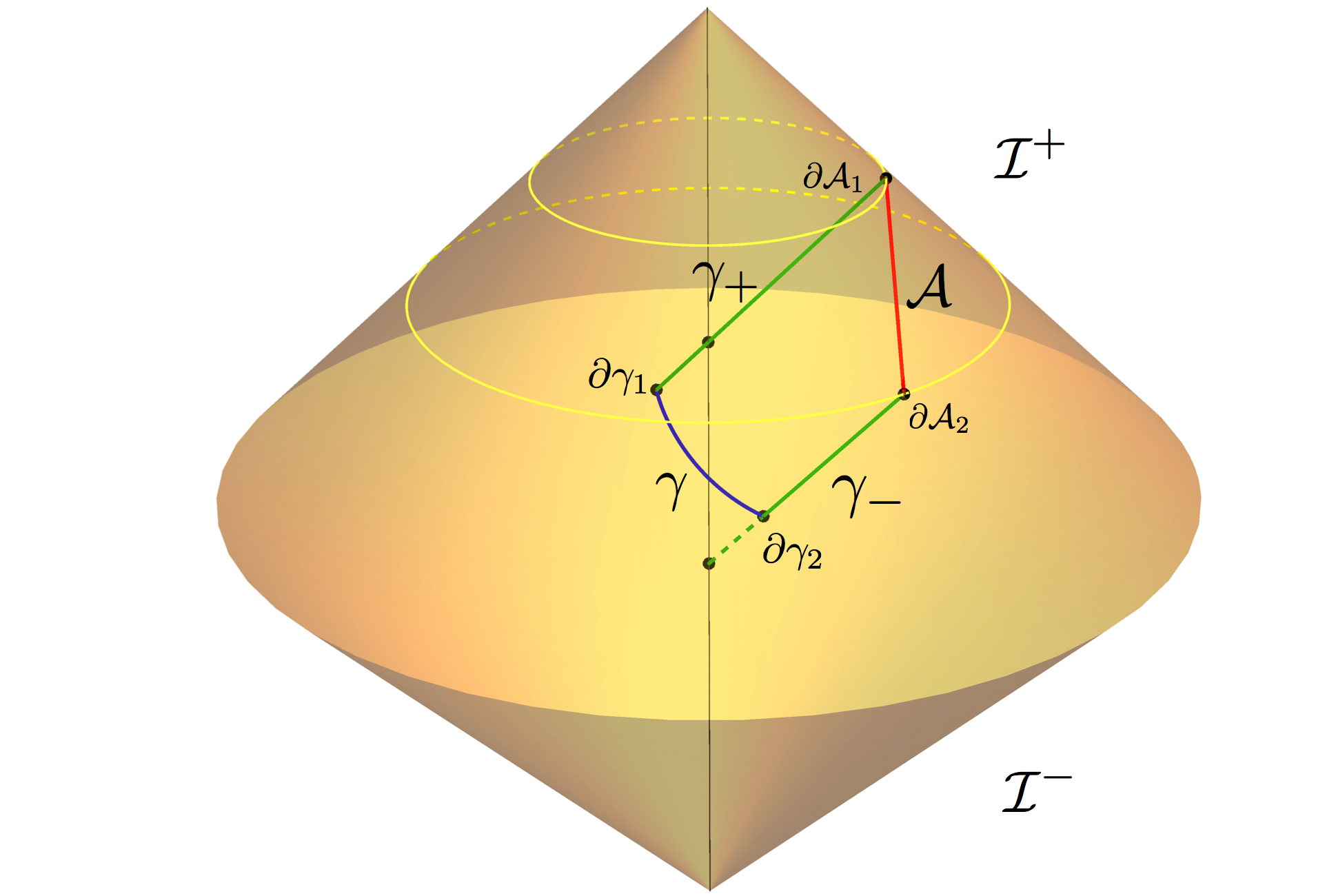} 
 \caption{The red line $\mathcal{A}$ on the future null infinity $\mathcal{I}^+$ is the boundary interval. 
 The blue line $\gamma$ is a spacelike geodesic, on which the bulk modular generator vanishes, $k_t^{\text{bulk}}=0$.
 The two green lines $\gamma_+$, $\gamma_-$ are null geodesics. The tangent vector on   $\gamma_+$ and  $\gamma_-$ are $k_t^{\text{bulk}}$.
The union  $\gamma_{\mathcal A}=\gamma\cup\gamma_+\cup\gamma_-$ is invariant under the modular flow $k_t^{\text{bulk}}$. The entanglement entropy is given by $S_{\text{HEE}}= {\text{Length}  (\gamma) \over 4G}={\text{Length}  (\gamma_{\mathcal A}) \over 4G}$.
\label{Penrose} }
\end{figure}

On the null surface $\mathcal{N}_+\cup \mathcal{N}_-$, there are three special curves $\gamma$ and $\gamma_{\pm}$ (see figure \ref{Penrose}). The spacelike curve $\gamma\subset \mathcal{N}_+\cap \mathcal{N}_-$ is a spacelike geodesic. 
Both the null curves $\gamma_\pm$ are along the $r$ direction. $\gamma_\pm$ starts from the end points $\partial A_{1,2}$ and intersect with the curve $\mathcal{N}_+\cap \mathcal{N}_-$ at $\partial\gamma_{1,2}$, respectively.

\subsection*{The spatial geodesic $\gamma$}
The curve $\gamma$ defined above is given by
\bea\label{gamma}
\gamma&=&\gamma_1\cup \gamma_2\,,\\
\gamma_1:\quad 
u&=&\frac{1}{2}  \cos \left(\frac{ l_\phi }{2}\right) \sqrt{4r^2-l_u^2 \csc ^2\left(\frac{l_\phi}{2}\right)}-r \,,
\nonumber\\
\phi&=& -\arcsin({l_u \csc{l_\phi\over2}\over 2r})\,,\nonumber\\
\gamma_2:\quad u&=&-\frac{1}{2}  \cos \left(\frac{ l_\phi }{2}\right) \sqrt{4r^2-l_u^2 \csc ^2\left(\frac{l_\phi}{2}\right)}-r \,,
\nonumber\\
\phi&=&\arcsin({l_u \csc{l_\phi\over2}\over 2r})-\pi\,,\nonumber\\
r_*&\le& r\le r_I, \quad r_*={l_u\over2}\csc {l_\phi\over2}, \quad r_I={l_u\over2} \csc^2 {l_\phi\over2}\,.\label{endpoint}
\eea
The spacelike geodesic $\gamma_1$ and $\gamma_2$ are smoothly connected at the ``turning point'' \be \gamma_1\cap \gamma_2:
\quad u_*=-{l_u\over2}\csc {l_\phi\over2},\,\phi_*=-{\pi\over2},\,r_*={l_u\over2}\csc {l_\phi\over2}, \quad \ee
At the maximum radius $r=r_I$, ${du\over dr}|_{r_I}=0$ the two end points are
\begin{align} \label{Bpoint}
 {\partial\gamma_1}:~(u_1,\phi_1,r_1)=&(\frac{l_u}{2}(1-2 \csc ^2\frac{l_{\phi }}{2}),{l_\phi\over2}-{\pi}\,,r_I)\,,
\\
\label{Apoint}
{\partial\gamma_2}:~(u_2,\phi_2,r_2)= &(-\frac{l_u}{2},-{l_\phi\over2}\,,r_I)\,.
\end{align} 
The end points (\ref{Apoint})(\ref{Bpoint}) can be obtained by looking at the image in $\widetilde{\text{FSC}}$ under \eqref{MinkBulk}. Note that $\tilde{M}$ and $\tilde{r}_c$ do not affect the null surfaces. 
For simplicity we set $\tilde{M}=1$ and $\tilde{r}_c=0$. We find the inverse coordinate transformation for \eqref{MinkBulk} is given by
\begin{align}
&r= \frac{1}{2}  \csc \frac{l_\phi}{2}
\times
\cr&\sqrt{l_u^2+4 \tilde{u}^2-4\tilde{r} \left(l_u \sin \frac{l_\phi}{2} \sinh (\tilde{\phi})+2 \tilde{u} \cos \frac{l_\phi}{2} \cosh (\tilde{\phi})+2 \tilde{u}\right)
+4 \tilde{r}^2 \left(\cos \frac{l_\phi}{2}+\cosh (\tilde{\phi})\right)^2}\,,
\cr
&u= \cot \frac{l_\phi}{2} (\tilde{r}-\tilde{u})- (r - \tilde{r}\csc \frac{l_\phi}{2}  \cosh (\tilde{\phi}))\,,
\cr
&\phi = -\cos ^{-1}\left[\frac{2 \left(\tilde{r} \cos \frac{l_\phi}{2} \cosh (\tilde{\phi})+\tilde{r}-\tilde{u}\right)}{2r \sin\frac{l_\phi}{2} }\right]\,.
\end{align}
The horizon $\tilde{r}=\tilde{r}_c=0$ of the $\widetilde{\text{FSC}}$ maps to the curve $\gamma$ \eqref{gamma}, with the range of $\gamma$ determined by the constraint $\tilde u\in[-{\Delta \tilde u\over2},\,{\Delta \tilde u\over2}]$ with $\Delta \tilde u$ given by \eqref{tuextensionGM}.


\subsection*{The null geodesics $\gamma_\pm$}
The null geodesic $\gamma_+$ is given by
\bea
\gamma_+\subset \mathcal{N}_+:~(u,\phi,r)=\Big\{  
\begin{array}{cc}
     (\frac{l_u}{2},\frac{l_\phi}{2},\tau_+),& \tau_+>0  \\ 
     (\frac{l_u}{2}+2\tau_+,\frac{l_\phi}{2}-\pi+\epsilon,-\tau_+)\,,&  -{l_u\over2} \csc^2 {l_\phi\over2}\le \tau_+\le 0  \\ 
  \end{array} 
\eea
while $\gamma_-$ is given by
\begin{align}
\gamma_-\subset \mathcal{N}_-:~(u,\phi,r)&=(-\frac{l_u}{2},-\frac{l_\phi}{2},r)\,,\quad r\ge {l_u\over2} \csc^2 {l_\phi\over2}\,.
\end{align}
It is easy to check that $\gamma_+$ connects the end points $\partial\mathcal{A}_1$ of the boundary interval and  $\partial\gamma_1$ of $\gamma$, while $\gamma_-$ connect the other two end points $\partial\gamma_2$ and $\partial\mathcal{A}_2$. The ranges of $r$ on $\gamma_\pm$ is determined by requiring that the coordinate transformation (\ref{phitilde}) is well defined, i.e. $
R_-\geq 0,~ R_+\geq 0\,,$ for the choice $\tilde {M}=1, \tilde J=0$. 
Both $\gamma_-$ and $\gamma_+$ maps to null curves (or just points for some choices of $\tilde M$ and $\tilde{r}_c$) on the horizon of $\widetilde{\text{FSC}}$. 

\subsection{The modular flow and its bulk extension}
Here we reconsider the physical meaning of the three geodesics $\gamma$ and $\gamma_{\pm}$ from the fixed points of the bulk modular flow $k_t^{\text{bulk}}$. We rewrite the modular flow \eqref{ktGMB} in the field theory side, and its bulk extension \eqref{ktGM}
\begin{align}\label{ktGM}
k_t=& \pi  \csc \left(\frac{l_{\phi }}{2}\right) \left(-l_u \csc \left(\frac{l_{\phi }}{2}\right)+l_u \cos (\phi ) \cot \left(\frac{l_{\phi }}{2}\right)+2 u \sin (\phi )\right)\partial_u
\cr
&+ 2 \pi \csc \left(\frac{l_{\phi }}{2}\right) \left(\cos \left(\frac{l_{\phi }}{2}\right)-\cos (\phi )\right) \partial_\phi\,,
\\
\label{ktGM}
k_t^{\text{bulk}} =& \pi  \csc \left(\frac{l_{\phi }}{2}\right) \left(-l_u \csc \left(\frac{l_{\phi }}{2}\right)+l_u \cos (\phi ) \cot \left(\frac{l_{\phi }}{2}\right)+2 u \sin (\phi )\right)\partial_u
\cr
 &+ \pi \csc \left(\frac{l_{\phi }}{2}\right)\Big(2\big(\cos \left(\frac{l_{\phi }}{2}\right)-\cos (\phi )\big) 
 + r^{-1}\left(l_u \sin (\phi ) \cot \left(\frac{l_{\phi }}{2}\right)
 -2 u \cos (\phi ) \right)
 \Big)\partial_\phi
 \cr
 &-\pi  \csc \left(\frac{l_{\phi }}{2}\right) \left(l_u \cos (\phi ) \cot \left(\frac{l_{\phi }}{2}\right)+2 (r+u) \sin (\phi )\right)\partial_r\,.
\end{align}
We see that $k_t$ vanishes at the boundary end points $\partial\mathcal{A}_{1,2}$. One can easily check that the fixed points which satisfy $k_t^{\text{bulk}}=0$ in the bulk is just the intersection curve $\mathcal{N}_+\cap\mathcal{N}_-$ (\ref{gamma}). 
Note that the curves $\mathcal{N}_+\cap\mathcal{N}_-$ (\ref{gamma}) can be extended even for $r>r_I$. However, if we take $r\rightarrow\infty$ in (\ref{gamma}), $\mathcal{N}_+\cap\mathcal{N}_-$ intersects with the boundary at $(u,\phi)=(-\pi,-\infty)$ and $(\pi,-\infty)$. 
The reason is that the boundary end points $\partial\mathcal{A}_{1,2}$ are the fixed points of $k_t$ but not the fixed points of $k_t^{\text{bulk}}$
\begin{align}
k_t|_{\partial\mathcal{A}_{1,2}}=0,\qquad k_t^{\text{bulk}}|_{\partial\mathcal{A}_{1,2}}=-2 \pi\left(r_I\pm r_{\infty}\right)\partial_r\neq 0\,,
\end{align}
and thus $\mathcal{N}_+\cap\mathcal{N}_-$ will never go through $\partial\mathcal{A}_{1,2}\,$. 
It is interesting that the orbits of $\partial\mathcal{A}_{1,2}$ under the bulk modular flow $k_t^{\text{bulk}}$ are just $\gamma_\pm$. 
 More interestingly, we find that $\gamma$ is the only extremal (or saddle) curve among all the curves that connect the null rays $\gamma_\pm$, the orbits of $\partial\mathcal{A}_{1,2}$ under bulk modular flow. This can be considered as a generalized version of the RT (HRT) proposal when the boundary entangling surface $\partial\mathcal{A}$ is not fixed under the bulk modular flow (or bulk extended replica symmetry).


\subsection{The geometric picture of holographic entanglement entropy}
We observe that the holographic entanglement entropy is given by the length of the spacelike geodesic $\gamma$ which is connected to the end points at the boundary by two null geodesics $\gamma_\pm$, 
\bea  S_{\text{\text{HEE}}}&=& {\text{Length} (\gamma) \over 4G}\eea

Alternatively, 
we can define a curve $\gamma_{\mathcal{A}}$ homologous to the boundary interval $\mathcal A$, \begin{align}\label{gammaA}
\gamma_{\mathcal{A}}=\gamma\cup \gamma_-\cup\gamma_+\,.
\end{align}
The holographic  entanglement entropy can also be given by the total length of  $\gamma_{\mathcal{A}}$
\bea S_{\text{HEE}} &=&{\text{Length}  (\gamma_{\mathcal A}) \over 4G}\,,\eea We hope to further clarify the choice between $\gamma$ and $\gamma_{\mathcal{A}}$ in the future.



\section{Holographic entanglement entropy from flat limit of AdS$_3$}\label{FlatAdS}
In this section, we re-derive the  HEE from a flat limit of AdS$_3$. As summarized in \cite{Casini:2011kv,Song:2016gtd}, the logic of Rindler method for AdS$_3$ is the following:
we do a quotient on AdS$_3$ and get a Rindler $\widetilde{\text{AdS}}_3$ black string, with the boundary of the later covers the a causal development of an interval on the boundary of the former. Accordingly the entanglement entropy of the interval equals to the Bekenstein-Hawking entropy of $\widetilde{\text{AdS}}_3$.

Under the flat limit $\ell\to \infty$, this picture provides a way to calculate the holographic entanglement entropy for BMSFT. The quotient on AdS$_{3}$ to Rindler $\widetilde{\text{AdS}}_3$ under the flat limit become a quotient on flat-space \eqref{fbtz} to Rindler $\widetilde{\text{FSC}}$  as described in section~\ref{GravityHEE}. Note that the outer horizon of the Rindler $\widetilde{\text{AdS}}_3$ is now pushed to an infinitely far away location, while the inner horizon becomes the Cauchy horizon of $\widetilde{\text{FSC}}$ . Thus, under the flat limit, the Bekenstein-Hawking entropy of the inner horizon of $\widetilde{\text{AdS}}_3$ is the quantity that gives the holographic entanglement entropy for the BMSFT \footnote{This differs from the prescription in \cite{Hosseini:2015uba}, which takes flat limit directly on the CFT results. This equals to taking flat limit on the Bekenstein-Hawking entropy of the outer horizon in $\widetilde{\text{AdS}}_3$.}.

In this section we first revisit the Rindler method for AdS$_3$ in section~\ref{6.1}. Then in section~\ref{6.2}, we change to Bondi gauge and discuss how to take flat limit. We re-derive the holographic entanglement entropy for BMSFT from a flat limit of AdS$_3$ for Einstein gravity in section~\ref{6.3} and TMG  in section~\ref{TMG} respectively.

\subsection{Holographic entanglement entropy for AdS$_3$ with Brown-Henneaux boundary conditions}\label{6.1}
\subsubsection*{Poincar\'{e} AdS$_3$}
For simplicity we first consider Poincar\'{e} AdS$_3$ \begin{align}\label{adsT0}
ds^2= \ell^2\Big(\frac{ d \rho^2}{4  \rho^2}+2  \rho d U d V\Big)\,.
\end{align}
The dual boundary field theory is a CFT$_{2}$ defined on a plane with zero temperatures.
Following the guide lines to construct Rindler transformations in section \ref{rindler}, we define $\partial_{\tilde{U}}$ and $\partial_{\tilde{V}}$ as a combination of global generators in either copy of the Virasoro algebras (see appendix B in \cite{Song:2016gtd} for an explicit example). Then we find the Rindler transformation for CFT$_2$ on a plane, and extend it to the bulk. The bulk transformation is a quotient on (\ref{adsT0}), and is given by
\begin{align}\label{ctUuads}
 T_{\tilde{U}} \tilde{U} =&\frac{1}{4}\log\Big[\frac{(1+ \rho(2 U+l_U) V)^2-  \rho^2 l_V^2(l_U/2+U)^2}{(1+ \rho(2U-l_U) V)^2-  \rho^2 l_V^2(l_U/2-U)^2}\Big]\,,
\cr
 T_{\tilde{V}} \tilde{V} =&\frac{1}{4}\log\Big[\frac{(1+ \rho(2 V+l_V) U)^2-  \rho^2 l_U^2(l_V/2+V)^2}{(1+ \rho(2V-l_V) U)^2-  \rho^2 l_U^2(l_V/2-V)^2}\Big]\,,
\cr
 \frac{\tilde{\rho}}{T_{\tilde{U}} T_{\tilde{V}}} =&\frac{\rho ^2 \left(l_U^2 \left(l_V^2-4 V^2\right)-4 U^2 l_V^2\right)+4 (2 \rho  U V+1)^2}{4 \rho  l_U l_V}\,.
\end{align}
Then we get the metric of the Rindler $\widetilde{\text{AdS}}_3$
\begin{align}\label{btztutv}
ds^2&=\ell^2\left(T_{\tilde{U}}^2 \,d \tilde{U} ^2+2  \tilde{\rho}   \,d \tilde{U}  d \tilde{V} + T_{\tilde{V}}^2 d \tilde{V} ^2 + \frac{d \tilde{\rho}  ^2 }{4 ( \tilde{\rho} ^2 -T_{\tilde{U}}^2 T_{\tilde{V}}^2)}\right)\,.
\end{align}
The asymptotic behavior of bulk quotient is given by
\begin{align}\label{adsRindler}
T_{\tilde{U}} \tilde{U}=&\text{ArcTanh}\left(\frac{2 U}{l_U}\right)+\mathcal{O}\left(\frac{1}{\rho}\right)\,,
\cr
T_{\tilde{V}} \tilde{V}=& \text{ArcTanh}\left(\frac{2 V}{l_V}\right)+\mathcal{O}\left(\frac{1}{\rho}\right)\,,
\end{align}
which is just the Rindler transformation and, as expected, a conformal mapping. It shows that the boundary of $\widetilde{\text{AdS}}_3$ (\ref{btztutv}) covers the causal development of an interval
\begin{align}
\mathcal{A}:\qquad (  - l_U/2 ,  - l_V/2  ) \rightarrow (  l_U/2 , l_V/2 )\,,
\end{align}
on the boundary of the original Poincar\'{e} AdS$_3$.
We introduce two infinitesimal parameters $\epsilon_U$ and $\epsilon_V$ and regulate the interval as
\begin{align}\label{intervalUV}
\mathcal{A}_{\text{reg}}:\qquad (  - l_U/2 +\epsilon_U,  - l_V/2+\epsilon_V  ) \rightarrow (  l_U/2-\epsilon_U , l_V/2-\epsilon_V ) \,.
\end{align}
These two parameters also regulate the extension of the $\tilde{U},~\tilde{V}$ coordinates of the $\widetilde{\text{AdS}}_3$,
\begin{align}
\Delta \tilde{U}=\frac{1}{T_{\tilde{U}}}\log\frac{l_U}{\epsilon_U}\,,
\qquad
\Delta \tilde{V}=\frac{1}{T_{\tilde{V}}}\log\frac{ l_V}{\epsilon_V}\,.
\end{align}
Hence the Bekenstein-Hawking entropy of the outer horizon of the $\widetilde{\text{AdS}}_3$ is given by
\begin{align}
S_{\text{outer}}=&\frac{\ell}{4G}\sqrt{T_{\tilde{U}}^2 \Delta \tilde{U}^2+2 T_{\tilde{U}} T_{\tilde{V}} \Delta \tilde{U} \Delta \tilde{V}+ T_{\tilde{V}}^2 \Delta \tilde{V}^2}
\cr
=&\frac{\ell}{4G}\Big|T_{\tilde{U}}\Delta \tilde{U}+T_{\tilde{V}} \Delta \tilde{V}\Big|
\cr
=&\frac{\ell}{4 G}\log\frac{l_U l_V}{\epsilon_U\epsilon_V}\,.
\end{align}
Following the logic of the Rindler method, $S_{\text{outer}}$ gives the holographic entanglement entropy for the single interval (\ref{intervalUV}) in a CFT$_{2}$ with zero temperature. This result is also consistent with the HRT \cite{Hubeny:2007xt} formula.

\subsubsection*{AdS$_3$ with a thermal (spatial) circle}
To consider CFT$_{2}$ with finite temperatures, we take the dual bulk spacetime to be a BTZ black string \footnote {Under the following transformations
\begin{align}\label{utot}
U\to \frac{\ell \varphi+t}{2 \ell^2}&,\quad V\to \frac{\ell \varphi-t}{2 \ell^2},\quad \rho \to 2 r^2-r_-^2-r_+^2\,,
\\
& T_{V}\to r_+-r_-,\quad T_U\to r_++r_-\,,
\end{align}
(\ref{adsT0}) can be written in the ADM formula
\begin{align}
ds^2=-\frac{\left(r^2-r_- ^2\right) \left(r^2-r_+ ^2\right)}{\ell^2 r^2}dt^2+\frac{\ell^2 r^2}{\left(r^2-r_- ^2\right) \left(r^2-r_+ ^2\right)}dr^2+r^2\left(d\varphi+\frac{ r_-  r_+ }{\ell r^2}dt\right)^2.
\end{align}}
\begin{align}\label{adsTg}
ds^2=\ell^2\left( T_{U}^2 dU^2+2\rho dU dV+T_{V}^2 dV^2+\frac{d\rho^2}{4(\rho^2-T_{U}^2 T_{V}^2)}\right),
\end{align}
with the thermal circle
\begin{align}\label{thermalcircleads}
\text{thermal circle}:~~(U,V)\sim\,(U+\frac{\pi i}{T_{U}},V-\frac{\pi i}{T_{V}})\,.
\end{align}
The mapping from Poincar\'{e} AdS$_3$ (\ref{adsT0}) to (\ref{adsTg}) is given by
\begin{align}\label{ctgeneralads}
U\to\,& e^{2 T_{U} U}\sqrt{1-\frac{2 T_{U} T_{V}}{\rho+T_{U} T_{V}}}\,,
\cr
V\to\,& e^{2 T_{V} V}\sqrt{1-\frac{2 T_{U} T_{V}}{\rho+T_{U} T_{V}}}\,,
\cr
\rho\to\,& \frac{(\rho+T_{U} T_{V})e^{-2(T_{U} U+T_{V} V)}}{4 T_{U} T_{V}}\,.
\end{align}
Accordingly, in terms of the coordinates of BTZ black string (\ref{adsTg}) we have
\begin{align}
\log \frac{l_U}{\epsilon_U}\to\log [\frac{\sinh \left( T_{U} l_U\right)}{T_{U} \epsilon_U}]\,, \qquad \log \frac{l_V}{\epsilon_V}\to\log [\frac{\sinh \left( T_{V} l_V\right)}{T_{V} \epsilon_V}]\,.
\end{align}
It is easy to see that the quotient on (\ref{adsTg}) to $\widetilde{\text{AdS}}_3$ (\ref{btztutv}) is just given by the combination of (\ref{ctUuads}) and (\ref{ctgeneralads}). Also the Bekenstein-Hawking entropy of the outer horizon of (\ref{btztutv}) changes to
\begin{align}
S_{\text{outer}}=\frac{\ell}{4 G}\log\left[\frac{\sinh \left( T_{U} l_U\right)\sinh \left( T_{V} l_V\right)}{T_{U} T_{V} \epsilon_U \epsilon_V}\right]\,.
\end{align}
This gives the holographic entanglement entropy for a single interval in a CFT$_{2}$ with finite temperatures.

When we consider imaginary temperatures $T_{U}\to -\frac{\pi i}{L_U}\,,T_{V}\to -\frac{\pi i}{L_{V}}$, the thermal circle (\ref{thermalcircleads}) changes to a spatial circle
\begin{align}\label{spatialcircleads}
\text{spatial circle}:~~(U,V)\sim\,(U-L_U,V+L_V)\,.
\end{align}
Correspondingly, the boundary field theory becomes a zero temperature CFT$_2$ on a cylinder.
Accordingly, the Bekenstein-Hawking entropy of the outer horizon of $\widetilde{\text{AdS}}_3$ (\ref{btztutv}) then becomes
\begin{align}
S_{\text{outer}}=\frac{\ell}{4 G}\log\left[\frac{L_U L_V\sin \left( \frac{\pi l_U}{L_U}\right)\sin \left( \frac{\pi l_V}{L_V}\right)}{\pi^2 \epsilon_U \epsilon_V}\right]\,.
\end{align}
As expected, we get the holographic entanglement entropy for a single interval in a CFT defined on the cylinder (\ref{spatialcircleads}).

\subsection{Flat limit of AdS$_3$}\label{6.2}
\subsubsection{Transformation to Bondi gauge and the flat limit}
It is more convenient to take the flat limit in the Bondi gauge.  Following \cite{Barnich:2012aw}, we find that, under the following coordinate transformations
\begin{align}\label{dipoleBMS}
U=& \frac{ \tanh ^{-1}\left(\frac{2 r}{T_V -T_U }\right)-\tanh ^{-1}\left(\frac{2 r}{T_V +T_U }\right)}{2  T_U }+\frac{\ell\phi+u}{2\ell^2}\,,
\cr
V=& \frac{ \tanh ^{-1}\left(\frac{2 r}{T_V -T_U }\right)+\tanh ^{-1}\left(\frac{2 r}{T_V +T_U }\right)}{2  T_V }+\frac{\ell\phi-u}{2\ell^2}\,,
\cr
\rho =& \frac{1}{2} \left(4 r^2-T_V ^2-T_U ^2\right)\,,
\end{align}
the BTZ black string (\ref{adsTg}) becomes
\begin{align}\label{adsbms}
ds^2=\left(8GM-\frac{r^2}{\ell^2}\right) du^2	-2 du dr+8GJdu d\phi+r^2d\phi^2\,,
\end{align}
with $M$ and $J$ defined by
\begin{align}\label{TUTVMJ}
T_U =& 2 \left(\sqrt{G \ell \left(\ell M-\sqrt{\ell^2 M^2-J^2}\right)}+\sqrt{G \ell \left(\ell M+\sqrt{\ell^2 M^2-J^2}\right)}\right)\,,
\cr
T_V =& 2 \left(\sqrt{G \ell \left(\ell M+\sqrt{\ell^2 M^2-J^2}\right)}-\sqrt{G \ell \left(\ell M-\sqrt{\ell^2 M^2-J^2}\right)}\right)\,.
\end{align}
In terms of the new coordinates $(u,\phi)$, the regulated interval (\ref{intervalUV}) now becomes
\begin{align}\label{intervalmuphi}
\mathcal{A}_{\text{reg}}:\qquad (  - l_u/2+\epsilon_u ,  - l_\phi/2 +\epsilon_\phi ) \rightarrow (  l_u/2 -\epsilon_u, l_\phi/2 -\epsilon_\phi)\,,
\end{align}
which is regulated by $\epsilon_{u}$ and $\epsilon_{\phi}$.
According to (\ref{dipoleBMS}), we find the relationships between the parameters
\begin{align}\label{lUlmu}
l_U=\frac{ \ell l_\phi+l_u }{2 \ell^2},\quad l_V= \frac{ \ell l_\phi-l_u}{2 \ell^2}\,,
\end{align}
\begin{align}\label{epsilons}
\epsilon_U=\frac{\epsilon_\phi}{2\ell}+\frac{\epsilon_u}{2\ell^2}\,,\qquad \epsilon_V=\frac{\epsilon_\phi}{2\ell}-\frac{\epsilon_u}{2\ell^2}\,.
\end{align}

We see that under the transformations \eqref{dipoleBMS} and \eqref{TUTVMJ}, the physical quantities in AdS$_3$ are functions of the physical quantities in Bondi gauge and the AdS radius $\ell$. The right way to take the flat limit is to take $\ell\to \infty$ while keeping all the physical quantities in Bondi gauge fixed.

\subsection{Holographic entanglement entropy in flat limit}\label{6.3}
\subsubsection*{Poincar\'{e} coordinate system}
The Bekenstein-Hawking entropy of the inner horizon $\rho=-T_{\tilde{U}} T_{\tilde{V}}$ of the Rindler $\widetilde{\text{AdS}}_3$ (\ref{btztutv}) is given by
\begin{align}\label{Sinnerbtz}
S_{\text{inner}}=&\frac{\ell}{4G}\Big|T_{\tilde{U}}\Delta \tilde{U}-T_{\tilde{V}} \Delta \tilde{V}\Big|
=\frac{\ell}{4 G}\left|\log\frac{l_U}{l_V}-\log\frac{\epsilon_U}{\epsilon_V}\right|
\cr
=&\frac{\ell}{4G}\left|\log\frac{\ell l_\phi+l_u}{\ell l_\phi-l_u}+\log\frac{\epsilon_V}{\epsilon_U}\right|\,,
\end{align}
where $l_U$ and $l_V$ depicts the interval on the boundary of the Poincar\'{e} AdS$_3$. Using (\ref{lUlmu}) and (\ref{epsilons}), we find the Bekenstein-Hawking entropy of the inner horizon under flat limit
\begin{align}\label{SHEET0}
S_{\text{\text{HEE}}}=S_C=S_{\text{inner}}|_{\text{flat limit}}=\frac{c_{\text{M}}}{6}\left|\frac{l_{u}}{l_{\phi}}-\frac{\epsilon_{u}}{\epsilon_{\phi}}\right|+\mathcal{O}\left(\frac{1}{\ell}\right)\,,
\end{align}
where $c_{\text{M}}=\frac{3}{G}$. This agrees with our previous result \eqref{Seeplane} with $c_L=0$.

\subsubsection*{FSC}
FSC can be considered as a flat limit of a BTZ solution. Note that under the flat limit, the outer horizon is pushed to  infinity. Thus the thermal circle in the bulk is actually associated to the inner horizon\footnote{The thermal circle associated to the outer horizon is given by $(U,V)\sim\,(U+\frac{\beta_{U} i}{\ell },V-\frac{\beta_{V} i}{\ell })$.}. In other words, we have
\begin{align}\label{thermalcircleinner}
\text{thermal circle}:~~(U,V)\sim\,(U+\frac{\beta_{U} i}{\ell },V+\frac{\beta_{V} i}{\ell })\,,
\end{align}
where $\beta_{U}$ and $\beta_{V}$ are finite quantities defined by $T_{U}=\frac{\pi \ell}{\beta_{U}}\,,T_{V}=\frac{\pi \ell}{\beta_{V}}$ . Then the thermal circle in the $(u,\phi)$ coordinates becomes
\begin{align}
\text{thermal circle}:~~(u,\phi)\sim \left(u+\beta_{u} i,\phi-\beta_{\phi} i\right)\,,
\end{align}
with $\beta_{u}=\ell(\beta_{U}-\beta_{V})\,, \beta_{\phi}=-(\beta_{U}+\beta_{V})$. Here we choose $\beta_{u}$ and $\beta_{\phi}$ as the physical quantities in Bondi gauge and keep them fixed when taking flat limit. Note that, the order of $\beta_U$ and $\beta_V$ in terms of $\ell$ are chosen such that we can get finite $\beta_u$ and $\beta_\phi$.

The Bekenstein-Hawking entropy of the inner horizon of the Rindler $\widetilde{\text{AdS}}_3$ is given by
\begin{align}
S_{\text{inner}}=\frac{\ell}{4G}\left|\log\frac{\beta_{U} \sinh(\pi\ell l_U/\beta_{U})}{\ell\epsilon_U}-\log\frac{\beta_{V} \sinh(\pi\ell l_V/\beta_{V})}{\ell\epsilon_V}\right|\,.
\end{align}
Then we use (\ref{lUlmu}) (\ref{epsilons}) and take the flat limit. We find
\begin{align}
S_{\text{\text{HEE}}}=\frac{c_{\text{M}}}{6}\left(\frac{\pi}{  \beta_{\phi}}\left(l_{u}+\frac{l_{\phi}\beta_{u}}{\beta_{\phi}}\right)\coth \frac{l_{\phi }\pi}{\beta_{\phi}}-\frac{\beta_{u}}{\beta_{\phi}}-\frac{\epsilon_{u}}{\epsilon_{\phi}}\right),
\end{align}
which agrees with \eqref{Seethermal} when $c_L=0$.

\subsubsection*{Global Minkowski}
Similarly when we consider imaginary temperatures $T_{U}\to -\frac{i\pi\ell}{L_U}\,,T_{V}\to -\frac{i\pi\ell}{L_{V}}$, the thermal circle (\ref{thermalcircleinner}) changes to a spatial circle
\begin{align}\label{spatialcircleflat}
\text{spatial circle}:~~(U,V)\sim\,(U-\frac{L_U}{\ell},V-\frac{L_V}{\ell})\,.
\end{align}
Compared with the finite temperatures case, this equals to replace $(\beta_{U},\beta_{V})$ with $(iL_{U},iL_{V})$ and replace $(\beta_{u},\beta_{\phi})$ with $(i L_{u},i L_{\phi})$. Accordingly the spatial circle in the $(u,\phi)$ coordinates is given by
\begin{align}
\text{spatial circle}:~~(u,\phi)\sim(u-L_{u},\phi+L_{\phi})\,.
\end{align}
The Bekenstein-Hawking entropy of the inner horizon of $\widetilde{\text{AdS}}_3$ (\ref{btztutv}) then becomes
\begin{align}
S_{\text{inner}}=\frac{\ell}{4 G}\left|\log\frac{L_U \sin \left( \pi l_U/L_U\right)}{\epsilon_U }-\log\frac{ L_V\sin \left( \pi l_V/L_V\right)}{ \epsilon_V}\right|\,,
\end{align}
which, after taking the flat limit, reduces to
\begin{align}
S_{\text{\text{HEE}}}=\frac{1}{2G}\left|\frac{\pi}{ L_{\phi}}\left( l_{u}+\frac{l_{\phi}L_{u}}{L_{\phi}}\right)\cot\frac{l_{\phi} \pi}{L_{\phi}}-\frac{L_{u}}{L_{\phi}}-\frac{\epsilon_{u}}{ \epsilon_{\phi}}\right|\,.
\end{align}
As expected, when we set $L_{\phi}=2\pi$ and $L_{u}=0$ the flat space is just global Minkowski. Then we have
\begin{align}
S_{\text{\text{HEE}}}=\frac{c_{\text{M}}}{6}\left|\frac{l_{u}}{2}\cot \frac{l_{\phi}}{2}-\frac{\epsilon_{u}}{ \epsilon_{\phi}}\right|\,,
\end{align}
which agrees with our previous result (\ref{Scylinder}) when $c_{\text{L}}=0$.

\section{Topologically massive gravity}\label{TMG}

\subsection{Topologically massive gravity in flat space} 

In the previous sections, we considered  the Einstein gravity and calculated  the holographic entanglement entropy. The results are in agreement with field theory calculations as well as other methods in literatures. However,  the asymptotic symmetry algebra of Einstein gravity has only one non-vanishing central charge $c_{\text{M}}$. In order to go beyond and incorporate the $c_{\text{L}}$ effects, we can consider the topologically massive gravity (TMG)~\cite{Deser:1981wh,Deser:1982vy}.

The action of TMG in AdS includes Einstein-Hilbert term, cosmological constant term and Chern-Simons term 
\be
\mathcal S_{\text{TMG}}= \frac{1}{16\pi G}\int d^3 x \;\sqrt{-g}
 \Bigg[ R+\frac{2}{\ell^2} +\frac{1}{2\mu}  \varepsilon^{\alpha\beta\gamma}  
\Big(\Gamma^{\rho}_{\,\alpha \sigma}   \partial_\beta\Gamma^\sigma_{\;\gamma\rho}
+\frac{2}{3}\Gamma^{\rho}_{\,\alpha \sigma} \Gamma^{\sigma}_{\; \beta\eta} \Gamma^{\eta}_{\;\gamma\rho} 
\Big)\Bigg]\,.
\ee
Einstein gravity is recovered in the limit $\mu\rightarrow \infty$.
The dual CFT is described by Virasoro algebra with left and right central charges \cite{Kraus:2005zm}
\be
c^+_{\text{TMG}}= \frac{3\ell}{2G}(1+\frac{1}{\mu \ell}),\quad c^-_{\text{TMG}}= \frac{3\ell}{2G}(1-\frac{1}{\mu \ell})\,.
\ee

In the flat limit 
\footnote{
Here we are simply choosing the flat limit  as $\ell \rightarrow \infty $.  A different  double scaling limit yields the so-called flat-space chiral gravity   \cite{Bagchi:2012yk} which is  argued  to be   unitary and ghost-free. For TMG in AdS, it generally admits ghost  fluctuations except at some critical point \cite{Li:2008dq}. Hence, the simple large $\ell$ limit of TMG is  expected to admit non-unitary fluctuations as well.  But  here  we assume that this is not  relevant for our entropy calculation in both field theory and gravity sides.  
}
$\ell\rightarrow \infty$, the cosmological constant disappears and we come to the TMG in flat spacetime. Asymptotic symmetry group analysis at null infinity yields the BMS$_3$ algebra \eqref{GCAcentral} with central charge~\cite{Bagchi:2012yk}
\be
c_{\text{L}}=\frac{3}{\mu G}, \qquad c_{\text{M}}=\frac{3}{G}\,.
\ee
Alternatively, these central charges  can be obtained from AdS by  taking Wigner-Inonu contraction: $c_{\text{L}}=c^+_{\text{TMG}} -c^-_{\text{TMG}} , c_{\text{M}}=(c^+_{\text{TMG}}+c^-_{\text{TMG}})/\ell $. 

The TMG also admits BTZ black hole solutions as in the Einstein gravity case
 \be
ds^2=-\frac{(r^2-r_+^2)(r^2-r_-^2)}{\ell^2 r^2} dt^2+ \frac{\ell^2 r^2} {(r^2-r_+^2)(r^2-r_-^2) } dr^2
+r^2\Big(d\varphi  - \frac{r_+r_-}{\ell r^2} dt \Big)^2
\ee
with
\be
\hat M=\frac{r_+^2+r_-^2}{\ell^2}, \quad  \hat J=\frac{2r_+r_-}{\ell}
\ee

But due to the presence of CS term, the physical conserved charges associated with BTZ in TMG get shifted \cite{Moussa:2003fc,Kraus:2005zm} 

\be
  \mathscr{M} =\hat M+\frac{\hat J}{\mu\ell^2}, \qquad   \mathscr{J}=\hat J+\frac{\hat M}{\mu}\,.
\ee

Fixing the physical charges $   \mathscr{M},  \mathscr{J}$ and taking  the flat limit  $\ell\rightarrow \infty$, we have 
\beqn
\hat M\rightarrow M=  \mathscr{M}, &\quad& \hat J \rightarrow J=   \mathscr{M}-\frac{  \mathscr{J}}{\mu}\,, \\
r_+ \rightarrow \ell \sqrt{  \mathscr{M}}=\ell \sqrt{M}, &\qquad&  
r_-\rightarrow \frac{  \mathscr{J}+  \mathscr{M}  /\mu}{2\sqrt{\hat{M}}} =\frac{J}{2\sqrt{M}} \equiv r_c \,,
\eeqn
and the BTZ metric reduces to the FSC metric after transforming to the Bondi gauge. 

The line element on the inner and outer horizon are
\beqn
ds^2_{\text{outer}}&=&(r_+ d\varphi +\frac{r_- dt}{\ell})^2  \xrightarrow{\ell \rightarrow \infty}  ( \sqrt{M} \ell d\varphi+r_c dt /\ell  ) ^2\,,\\
ds^2_{\text{inner}}&=&(r_- d\varphi +\frac{r_+dt}{\ell})^2   \xrightarrow{\ell \rightarrow \infty}   ( r_c d\varphi+\sqrt{M} dt)^2\,.
\eeqn

\subsection{Thermal entropy formula from flat limit} 

 The thermal entropy associated with the inner and outer horizons of the BTZ black hole  can be obtained as \cite{Solodukhin:2005ah,Sahoo:2006vz,Park:2006gt,Tachikawa:2006sz}
\be
S_{\text{inner}}=\frac{\ell_{\text{inner horizon}}}{4G}+\frac{\ell_{\text{outer horizon}}}{4G \mu\ell} ,\qquad
S_{\text{outer}}=\frac{\ell_{\text{outer horizon}}}{4G}+\frac{\ell_{\text{inner horizon}}}{4G \mu\ell}\,.
\ee
where the length of horizon comes from integration. 

 
 In the flat limit, the thermal entropy of the $ {\text{FSC}}$ comes from the inner horizon of BTZ black hole, so the entropy should be
 \beqn\label{Sftmg}
S_{\text{$ {\text{FSC}}$  TMG}}&=&S_{\text{inner}}|_{\ell \rightarrow \infty}
=
 \frac{\sqrt{M}\Delta t+r_c \Delta \varphi}{4G} +\frac{\sqrt{M} \ell \Delta \varphi }{4G\mu \ell} 
\\&=&
\frac{\sqrt{M}\Delta {  u}+r_c \Delta {  \phi}}{4G} +\frac{\sqrt{M}\Delta {  \phi}}{4G\mu} 
\equiv S_{\text{BH}}+S_{\text{CS}}\,,
 \eeqn
where in the second line, we have transformed to Bondi gauge. The CS correction to thermal entropy is 
\begin{align}\label{CSEE}
S_{\text{CS}}=\frac{\sqrt{M}\Delta {\phi}}{4G\mu}\,. 
\end{align}

\subsection{Thermal entropy formula from direct calculation}

In the previous subsection, we derived the contribution of Chern-Simons term to the entropy by taking the flat limit of BTZ black hole. This approach is physically more intuitive. However, considering the possible subtlety of flat limit, a more direct calculation without involving BTZ is also very desirable. This can be done by performing a symplectic analysis. As elaborated in \cite{Tachikawa:2006sz}, the shift of thermal entropy due to the Chern-Simons term in 3D can be calculated from the following formula:
\be
S_{\text{CS}}= \frac{1}{4  G\mu} \int_\Sigma \Gamma_N,   \qquad \text{with } \quad 
\Gamma_N=-\frac{1}{2} \epsilon^{\mu}_{\;\;\sigma} \Gamma^{\sigma}_{\mu\rho} dx^\rho\,,
\ee
where  the bifurcation horizon $\Sigma$ is a co-dimension 2 surface at radius $ r=r_c\equiv J/2\sqrt{M} $,
generated by the Killing vector 
\be
\zeta= \partial_u -\frac{2M}{J} \partial_\phi\,.
\ee

The binormal vector $\epsilon^{\mu\nu}$ is defined as 
\be
\epsilon^{\mu\nu}=n^{\mu} \zeta^{\nu} - n^{\nu} \zeta^{\mu}\,,
\ee
 where 
  $n^{\mu}$ is  null vector normal to $\Sigma$, given by 
\be
n =\partial_r\,.
\ee

Then, we can easily find that $\epsilon^{ur}=-\epsilon^{ru}=-1, \epsilon^{r\phi}=-\epsilon^{\phi r}=-\frac{2M}{J}$ and  
\be
\epsilon^{\mu}_{\;\;\sigma} \Gamma^{\sigma}_{\mu u} = \epsilon^{\mu}_{\;\;\sigma} \Gamma^{\sigma}_{\mu r} =0,
\quad
\epsilon^{\mu}_{\;\;\sigma} \Gamma^{\sigma}_{\mu\phi} =-2\sqrt{M}  \,.
\ee

Therefore, 
\be
S_{\text{CS}}= \frac{1}{4  G\mu} \int_\Sigma \frac{-1}{2} (-2\sqrt{M})d\phi=\frac{\sqrt{M}\Delta\phi}{4 G\mu}\,,
\ee
where the $\Delta\phi$ is the extension of $\Sigma$ along $\phi$ direction.  This agrees with the previous results \eqref{CSEE} obtained by taking flat limit.

\subsection{CS contribution to holographic entanglement entropy}

In previous sections, we have derived the holographic  entanglement entropy in Einstein gravity by considering Bekenstein-Hawking entropy in the Rindler spacetime. As expected,  it reproduces the results in   dual field theory with $c_{\text{L}}=0$. Now  by considering the TMG, we get a non-vanishing central charge $c_{\text{L}}$ in the dual field theory. The effects of $c_{\text{L}}$ on holographic  entanglement entropy can be similarly calculated by employing the CS correction to the thermal entropy $S_{\text{CS}}$.  

Using Rindler method, the contribution of CS term to entanglement entropy boils down to the quantity $\Delta\tilde\phi$ in  Rindler spacetime $\widetilde{\text{FSC}}$. Since we have calculated $\Delta \tilde{\phi}$ in different cases (see \eqref{noextension}, \eqref{fscextension}, \eqref{gmextension}), it is straightforward to get entanglement entropy corrections due to the CS term
 \begin{align}
\text{ Poincar\'{e}:}&~~S_{\text{CS}}=\frac{1}{2\mu G} \log \frac{l_\phi}{\epsilon_\phi}\,,
\\
\text{ FSC:}&~~S_{\text{CS}}=\frac{1}{2\mu G}\log \frac{2\sinh{\frac{\sqrt{M}l_\phi}{2}}}{ \sqrt{  M} \epsilon_\phi}\,,
\\
\text{Global Minkowski:}&~~S_{\text{CS}}=\frac{1}{2\mu G} \log \frac{2\sin {\frac{ l_\phi}{2}}}{ \epsilon_\phi}\,.
 \end{align}
Together with the Bekenstein-Hawking part considered previously, it is easy to check that the total $S_{\text{\text{HEE}}}$ exactly agrees with field theory results.

\section{R\'{e}nyi  entropy}\label{Renyientropy}

In this section, we will   use Rindler method to derive the R\'{e}nyi  entropy. Under the Rindler transformation, the density matrix is transformed unitarily. Thus,   the entanglement entropy and R\'{e}nyi  entropy in the original  space are the same as the thermal entropy and thermal Reyni entropy in the Rindler space respectively. Therefore, we only need to consider the thermal R\'{e}nyi  entropy in the Rindler space and make a connection between thermal R\'{e}nyi  entropy and thermal entropy.   Then, the relation  between  R\'{e}nyi  entropy and  entanglement entropy in the original space follows directly. 

\subsection*{Field Theory Side}

The thermal R\'{e}nyi  entropy  on arbitrary torus is defined as
\be
S_ {\bar{b}|b}^{(n)}({\bar{a}|a} ) =\frac{1}{1-n} \log \Big( \frac{Z_ {\bar{b}|b}(n{\bar{a}|na} )}{Z_ {\bar{b}|b}({\bar{a}|a} )^n}  \Big)\,.
\ee
By performing a BMS coordinate transformation, the R\'{e}nyi  entropy on arbitrary torus is the same as the  R\'{e}nyi  entropy on the canonical torus
\be
S_ {\bar{b}|b}^{(n)}({\bar{a}|a} )=S_{0|1}^{(n)}(\hat\beta_u | \hat\beta_\phi ) =\frac{1}{1-n} \log \Big( \frac{Z_ {0|1}(n\hat\beta_u,n\hat\beta_\phi)}{Z_ {0|1}(\hat\beta_u,\hat\beta_\phi)^n} \Big)\,,
\ee
where the partition on the canonical torus, as justified in appendix \ref{CardyThermal}, can be approximated by
\be
Z_{0|1} ( \hat\beta_u|\hat\beta_\phi ) \approx\exp\Big(-  c_{\text{M}}   \frac{\pi^2\hat\beta_u}{6\hat\beta_\phi^2}  - c_{\text{L}} \frac{ \pi^2}{6 \hat\beta_\phi} \Big)\,.
 \ee
Thus,
 \be
S_ {\bar{b}|b}^{(n)}({\bar{a}|a} ) =     - \frac{\pi^2}{6} (1+\frac{1}{n})  \Big( \frac{ c_{\text{L}} }{\hat\beta_\phi}+ \frac{  c_{\text{M}}\hat\beta_u   }{\beta_\phi^2}  \Big)     =\frac{1}{2}(1+\frac{1}{ n})S^{(1)}_ {\bar{b}|b} ({\bar{a}|a} ) \equiv \frac{1}{2}(1+\frac{1}{ n})S_ {\bar{b}|b} ({\bar{a}|a} )\,.
\ee
 
Therefore, for all the cases we considered before, one can obtain   the following relation  between R\'{e}nyi  entropy and entanglement entropy by using the Rindler method
\bea
S^{(n)}_{\text{BMSFT}}=\frac{1}{2}(1+\frac{1}{ n})S_{\text{EE}}\,.
\eea

Obviously, the   R\'{e}nyi  entropy is alway proportional  to entanglement entropy by a fixed coefficient and reduces to the entanglement entropy in the limit $n\rightarrow 1$. In addition, the coefficient in BMSFT in the same as that in CFT. This is expected from the fact that BMSFT is a specific limit of CFT.

 \subsection*{Gravity Side}

Consider the periodic identified FSC $\tilde \phi\sim \tilde \phi+2\pi $
\footnote{For general periodic identification, one can discuss similarly by calculating the conserved charges more carefully. Or one can consider a further linear coordinate transformation to arrive at the torus $\hat \phi \sim\hat \phi+2\pi$.}, 
the relevant thermal dynamic quantities  associated with the Cauchy horizon, including thermal entropy, Hawking temperature and angular velocity, are given by\footnote{Here we restore the $G$-dependence of all the quantities, especially $M,J$ in the previous sections should be $M,J\rightarrow 8GM, 8GJ$.}
\be
S_C= \frac{J \pi }{\sqrt{2GM}}  +\frac{\sqrt{2GM}\pi}{G\mu}, \quad T_C=\frac{2M \sqrt{2GM} }{\pi J},
\quad \Omega_C=\frac{2M}{J}\,,
\ee
where the second term in $S_C$ has been derived in section \ref{TMG}.
The physical conserved charges are
\be
  \mathscr{M}=M, \quad   \mathscr{J}=J+M/\mu\,.
\ee
 Then, it is easy to show that the system satisfies the first law \cite{Barnich:2012xq}
 \be
 d \mathscr{M}=-T_C dS_C+\Omega_C d \mathscr{J}\,.
 \ee
 Note that the first law is not the conventional one due to the minus sign before $T_C$.
 The partition function is thus defined as
 \be
 \ln Z_C(\beta_C, \Phi_C) \equiv-S_C-\beta_C \mathscr{M}+\Phi_C \mathscr{J}\,,
 \ee
 where $\beta_C=T_C^{-1}, \Phi_C=\beta_C\Omega_C$.

 One can show that
 \be
 \ln Z_C(\beta_C, \Phi_C) =-\frac{\pi^2}{2G} \Big( \frac{\beta_C}{\Phi_C^2}+\frac{1}{\mu \Phi_C} \Big)\,.
 \ee
 The R\'{e}nyi  entropy is defined as
 \be
 -S^{(n)}=\frac{1}{1-n} \log \Tr \rho^n\,,
 \ee
where we put an additional minus sign to take account of the unconventional first law, otherwise the entropy obtained below is negative. The normalized density matrix is
 \be
 \rho=\frac{ \exp\Big[ -\beta_C\mathscr{M}+\Phi_C\mathscr{J}  
 \Big] }{  \Tr \exp\Big[ -\beta_C\mathscr{M}+\Phi_C\mathscr{J} 
 \Big]}\,.
 \ee
Then, we can obtain the  the thermal R\'{e}nyi  entropy 
  \be 
 -S^{(n)}_{\text{bulk}}= \frac{1}{1-n} \log  \frac{Z_C(n\beta_C, n\Phi_C) }{Z_C(\beta_C, \Phi_C) ^n}
 =  (1+\frac{1}{ n}) \ln Z_C(\beta_C, \Phi_C)\,.
 \ee

This yields the following relations
 \be\label{RenyiThermal}
 S^{(n)}_{\text{bulk}}=\frac{1}{2}(1+\frac{1}{ n}) S^{(1)}_{\text{bulk}} \equiv \frac{1}{2}(1+\frac{1}{ n}) S_{\text{thermal}} \,,
 \ee
 where the thermal entropies include the Bekenstein-Hawking term and the Chern-Simons term contribution.
 
After performing a   Rindler transformation, \eqref{RenyiThermal} is translated to the relations between holographic R\'{e}nyi entropy(HRE) and holographic entanglement entropy(HEE)
 \be
 S^{(n)}_{\text{HRE}} = \frac{1}{2}(1+\frac{1}{ n}) S_{\text{\text{HEE}}}\,,
 \ee 
 
Comparing the results of gravity and field theory, we can find that the R\'{e}nyi  entropy agrees on both sides
 \be  S^{(n)}_{\text{HRE}}= S^{(n)}_{\text{BMSFT}}\,.\ee

\section*{Acknowledgement}
We thank  for helpful discussions with  A.~Castro, C.~Chang,  G.~Comp\`{e}re, B.~Czech,  M.~Guica, T.~Hartman, R.~Miao, R.~Myers, M.~Rangamani, and J.~Xu. This work was supported in part by start-up funding from Tsinghua University. W.S. is also supported by the National Thousand-Young-Talents Program of China. H.J. is supported by grants HKUST4/CRF/13G and ECS 26300316 issued by the Research Grants Council (RGC) of Hong Kong. H.J. would like to thank the Yau Mathematical Sciences Center in Tsinghua University for kind hospitality.  

\appendix

\section{``Cardy'' formula in BMSFT revisited}\label{CardyThermal}
Consider a BMSFT living on an arbitrary  torus with the following identification
\be
(\tilde u,\tilde \phi ) \sim (\tilde u+i \bar{a},\tilde \phi-i a )
\sim (\tilde u+2\pi \bar{b},\tilde \phi-2\pi b)\,.
\ee
The partition function of BMSFT on such torus is defined as
\be
Z_{\bar{b}|b} ({\bar{a}|a}) \equiv \Tr_{\bar{b}|b} \Big(e^{- \bar{a} {\mathcal M}_0^{\bar{b}|b} } e^{ a {\mathcal L}_0^{\bar{b}|b} } \Big) \,,
\ee
where in the $a$-circle is viewed as the thermal circle, and
 ${\mathcal M}_0^{\bar{b}|b}, {\mathcal L}_0^{\bar{b}|b} $ are the charges  generating the translation along $u$ and $\phi$ directions, defined on the spatial $b$-circle.
 
The spectrum of a field theory is usually discussed on certain canonical circle which is a $2\pi$ spatial circle along $\phi$. Under the BMS transformation
\be
\hat  \phi=\frac{\tilde \phi }{b}, \qquad  \hat u=\frac{\tilde u }{b}  +\frac{\bar{b}}{b^2}\tilde \phi\,, \label{canonical}
\ee
the new torus has a canonical spatial circle,
\bea
(\hat u,\hat  \phi)  &\sim& ( \hat u+i\hat\beta_u,\hat \phi-i\hat\beta_\phi)\sim (\hat u,\hat \phi-2\pi)\,,
\\ \label{ModularParameter}
\hat\beta_\phi&=&\frac{ a}{b} , \quad \hat\beta_u=\frac{  \bar{a}b  -a \bar{b} }{b^2}\,.
\eea
Note that the Schwarzian derivative for the BMS transformation (\ref{canonical}) vanishes, thus, the new charges from (\ref{chargetr}) will not acquire any anomalous terms, and the partition function is invariant under this transformation
\bea\hat
Z_ {\bar{b}|b}({\bar{a}|a} )&=&Z_{0|1} ({\hat\beta_u |\hat\beta_\phi})\,,\nonumber\\&=&\Tr_{0|1} \Big(e^{- \hat\beta_u {\mathcal M_0}^{\text{Cyl}} } e^{\hat\beta_\phi {\mathcal L_0}^{\text{Cyl}}  } \Big)\,,\nonumber\\
&=&\Tr_{0|1} \Big(e^{-\hat\beta_u (\mathcal M_0-\frac{c_{\text{M}} }{24}) } e^{\hat\beta_\phi (\mathcal L_0-\frac{c_{\text{L}}}{24}  )} \Big)\,,\label{zlm}
\eea
where the ${\mathcal M_0}^{\text{Cyl}},{\mathcal L_0}^{\text{Cyl}}  $ are the charges  of $\p_u$ and $\p_\phi$ on the cylinder with the canonical spatial cycle, and on the third line we have used \eqref{PlanCylLM} to express the partition function in terms of the charges on the plane.

Next we perform the following BMS transformation (or $S$-transformation) that exchanges the spatial and thermal circle of the torus,
\begin{align}\label{exchange}
\hat{\phi}'=2\pi i\frac{\hat{\phi}}{\hat\beta_\phi} ,\qquad \hat{u}'=\frac{2\pi i}{\hat\beta_\phi}(\hat{u}+\frac{\hat\beta_u}{\hat\beta_\phi} \hat{\phi}).
\end{align}
The torus under this transformation become
\begin{align}
(\hat{u}',\hat{\phi}')\sim(\hat{u}',\hat{\phi}'+2\pi)\sim(\hat{u}'+i\hat\beta'_u,\hat{\phi}'-i\hat\beta'_\phi)\,,
\end{align}
with
\begin{align}
\hat\beta'_u=-4\pi^2\frac{ \hat\beta_u}{\hat\beta_\phi^2}\,,\qquad \hat\beta'_\phi=\frac{4\pi^2}{\hat\beta_\phi}\,.
\end{align}
Again, the Schwarzian derivative for the BMS transformation (\ref{exchange}) vanishes, thus, no anomalous terms appear and we have 
\bea
Z_{0|1} ( \hat\beta_u|\hat\beta_\phi )&=&Z_{0|1} ( -4\pi^2\frac{ \hat\beta_u}{\hat\beta_\phi^2} |\frac{4\pi^2}{\hat\beta_\phi} )\,.\label{ModInv}
\eea
This can be regarded as the modular invariance of BMSFT. This modular invariance agrees with the modular invariance inherited from CFT$_2$ under flat limit \cite{Bagchi:2012xr,Barnich:2012xq}.

For a CFT, the Cardy formula is valid in certain parameter region \cite{Cardy:1986ie,Hartman:2014oaa} where the partition function can be approximated by the ground state contribution after the S-transformation.
Similarly, in the ``Cardy region'' of GCFT, vacuum contribution is expected to dominate the partition function, so that
\beqn\label{PartFunZ}
Z_{0|1} ( \hat\beta_u|\hat\beta_\phi )=Z_{0|1} ( -4\pi^2\frac{ \hat\beta_u}{\hat\beta_\phi^2} |\frac{4\pi^2}{\hat\beta_\phi} )   &=&
\Tr \Big(e^{  4\pi^2\frac{ \hat\beta_u}{\hat\beta_\phi^2} (\mathcal M_0-\frac{c_{\text{M}} }{24})  }  e^{\frac{4\pi^2}{\hat\beta_\phi} (\mathcal L_0-\frac{c_{\text{L}}}{24})   }
\Big)\,,
 \nonumber \\
&\approx& \exp\Big(-  c_{\text{M}}   \frac{\pi^2\hat\beta_u}{6\hat\beta_\phi^2}  - c_{\text{L}} \frac{ \pi^2}{6 \hat\beta_\phi} \Big)\,,
\eeqn
where the vacuum charges on the plane have been taken to be zero.
 In this paper, we will not attempt to give a necessary condition for (\ref{PartFunZ}), due to some unusual properties of BMSFT including the abnormal first law of thermodynamics \cite{Bagchi:2012xr,Barnich:2012xq} and the generic non-unitarity of the highest weight representation \cite{Bagchi:2009pe}. A sufficient condition for the last line of (\ref{PartFunZ}) is that the charges $\mathcal L_0, \mathcal M_0$  are bounded from below, and 
 \be \hat\beta_\phi \rightarrow 0^-, \qquad \hat\beta_u/\hat\beta_\phi^2 \rightarrow \mathbb{R}^-  \,. \label{cardyregion}\ee
Then the thermal entropy is approximately
\bea
S_ {\bar{b}|b}({\bar{a}|a} )&=&S_{0|1}(\hat\beta_u| \hat\beta_\phi)=(1-\hat\beta_u \partial_{\hat\beta_u}-\hat\beta_\phi \partial_{\hat\beta_\phi} )\log Z_{0|1} (\hat\beta_u|\ \hat\beta_\phi )\,,\\
& =&- \frac{\pi^2}{3}  \Big( \frac{ c_{\text{L}} }{\hat\beta_\phi}
+ \frac{  c_{\text{M}}\hat\beta_u   }{\hat\beta_\phi^2}  \Big)\,,\\
&=& - \frac{\pi^2}{3}  \Big(c_{\text{L}} \frac{ b  }{a}
+ c_{\text{M}}\frac{  (\bar a b-a \bar b)  }{a^2}  \Big)\,.
\eea

\section{Killing vectors}\label{KillingVec}
We focus on the local isometries of the solutions \eqref{ClassSol}, which can be obtained from the Killing equations $\Lie_\xi g_{\mu\nu}=0$. The general solutions have the following form
\beqn\label{BlkKilling}
 \xi^u &=&  u \partial_\phi Y(\phi)+T(\phi)\,,  \\
 \xi^\phi &=& Y(\phi) -\frac{u}{r} \partial_\phi\partial_\phi Y(\phi)-\frac{1}{r} \partial_\phi T(\phi)\,,\\
 \xi^r&=&-\frac{J}{2r} \partial_\phi \xi^u-r \partial_\phi \xi^\phi\,,
 \eeqn
 where $Y$ and $T$ satisfy the following equations
 \beqn
 M \partial_\phi Y-\partial_\phi^3 Y &=&0\,, \\
 J\partial_\phi Y +M \partial_\phi T-\partial_\phi^3 T  &=&0\,.
 \eeqn

A general Killing vector can be obtained and written in the following form
\be
\xi=\sum_{i=-1}^1b_i L_i+d_i M_i\,,
\ee
where $b_i,c_i$ are arbitrary constants and $L_i,\,M_i$ are the normalized to satisfy  the sub-algebra of the GCA  
 \besub
\beqn
\big[  L_i,L_j]  &=&  (i-j)L_{i+j}\,,\\
\big[L_i,M_j]  &=&  (i-j)M_{i+j} \,,\qquad i,j=\{-1,0,1\}\,,\\
\big[ M_i,M_j]  &=&  0\,,
\eeqn
\eesub
The explicit form of the generators depend on the background, which will be displayed below.
\subsection{FSC}
 
The most general solutions can be found
\beqn
Y(\phi)&=&\sum_{j=-1}^1 {b_j \over \sqrt{M}} e^{-j \sqrt{M}\phi},\quad 
 \\
 T(\phi) &=& \sum_{j=-1}^1 \Big(b_j {J\over 2 M^{3\over2}}(\phi\p_\phi -1)e^{-j\sqrt{M}\phi}-{d_j \over \sqrt{M}} e^{-j\sqrt{M}\phi}\Big)\,.
 \eeqn
More explicitly, the Killing vectors are ( $\xi=\xi^u\partial_u+\xi^r\partial_r+\xi^\phi\partial_\phi \equiv (\xi^u,\xi^r,\xi^\phi)$ )

\begin{subequations}\label{KillingFSC}
\beqn
L_1&=&-  e^{-\sqrt{M} \phi } \Bigg( \frac{  (\sqrt{M} r_c \phi +M u+r_c)}{M},\frac{ (r+r_c) (\sqrt{M} r_c \phi +M u-r)}{r}
, \frac{ (\sqrt{M} r_c \phi +M u-r)}{\sqrt{M} r} \Bigg)\,, \qquad\quad
\\
L_0&=&\Big(-\frac{r_c}{M},0,\frac{1}{\sqrt{M}}\Big)\,,
\\
L_{-1}&=&e^{\sqrt{M} \phi }\Bigg(  \frac{  (\sqrt{M} r_c \phi +M u-r_c)}{M}, \frac{  (r-r_c) ( \sqrt{M} r_c \phi +M u-r)}{r},-\frac{(\sqrt{M} r_c \phi +M u-r)}{\sqrt{M} r}  \Bigg)\,,
\\
M_1&=&-  e^{-\sqrt{M} \phi } \Big(\frac{1}{\sqrt{M}}, \frac{\sqrt{M}   (r+r_c)}{r}, \frac{1}{r}\Big)\,,
\\
M_0&=&\Big(-\frac{1}{\sqrt{M}},0,0\Big)\,,
\\
M_{-1}&=&e^{\sqrt{M} \phi } \Big(-\frac{1}{\sqrt{M}},-\frac{\sqrt{M}   (r-r_c)}{r},\frac{1}{r}\Big)\,.
\eeqn
\end{subequations}

%
%
%
%

\subsection{Poincar\'{e} patch}
In the Poincar\'{e} patch, $M=J=0$,  the Killing vectors are \begin{align}\label{NOKilling}
L_1&=(-2u\phi,\; 2r\phi, \; \frac{2u}{r}-{\phi^2}) \; ,       &   L_0&= (-u,\;  r,  \;    -\phi)    \; ,            &  L_{-1}&=(0,0,-1) \;,   \nonumber  \\
M_1&= ({\phi^2},    \;2,     \;-\frac{2\phi}{r})     \; ,          &      M_0&=(\phi,\; 0,\;   -\frac{1}{r})  \;,  &  M_{-1}&=(1,0,0) \; .
\end{align}

One can even show that the above Killing vectors can also be obtained by taking the flat limit of the  $\text{SL}(2,\mathbb R)\times \text{SL}(2,\mathbb R)$ algebra of Poincar\'{e} AdS: expressing the AdS Killing vectors in terms of BMS coordinate and then taking Wigner-Inonu contraction.

\subsection{Global Minkowski}

This is the global Minkowski spacetime and the isometry group is the Poincar\'{e} group which is composed of translation, spatial-rotation and boost.  
\besub\label{KillingGM}
\beqn
L_1&=&-  e^{-i \phi } \Bigg( u,-u-r
, i ({u\over r}+1) \Bigg)\,, \qquad\quad
\\
L_0&=&\Big(0,0,-i\Big)\,,
\\
L_{-1}&=&e^{i \phi }\Bigg( u, -u-r,-i({u\over r}+1) \Bigg)\,,
\\
M_1&=&e^{-i \phi } \Big(i, -i , -\frac{1}{r}\Big)\,,
\\
M_0&=&\Big(i,0,0\Big)\,,
\\
M_{-1}&=&e^{i \phi } \Big(i,-i, \frac{1}{r}\Big)\,.
\eeqn
\eesub

\section{Rindler transformations for BMSFT }\label{RindlerTsFSC}

We follow the procedure in section~\ref{rindlerTranf} to derive the Rindler transformation for BMSFT with a thermal or spatial circle. 

\subsection{Thermal BMSFT} \label{FSCTs}

For finite temperature BMSFT dual to FSC, the symmetry generators that preserves vacuum can be obtained from the bulk Killing vectors \eqref{KillingFSC} by   keeping  the $u, \phi$  components and then taking limit $r\rightarrow \infty$.  By matching with the general form of generators \eqref{FieldKillingForm}, we can get the following two functions  
\beqn
Y&=&\frac{\beta_\phi}{2\pi}\Big( b_0+b_1 e^{-\frac{2\pi \phi}{\beta_\phi}} + b_{-1} e^{ \frac{2\pi \phi}{\beta_\phi}}  \Big)\,, \\
T&=&- \frac{1}{2\pi} \Big(    
 e^{ \frac{2\pi \phi}{\beta_\phi}}  (b_{-1}\beta_u +d_{-1}\beta_\phi-\frac{2\pi b_{-1} \phi \beta_u}{\beta_\phi})
 +b_0\beta_u+d_0\beta_\phi
\nonumber \\&&\qquad 
+e^{-\frac{2\pi \phi}{\beta_\phi}}  (b_{1}\beta_u +d_{1}\beta_\phi+\frac{2\pi b_1 \phi \beta_u}{\beta_\phi})
\Big)    \,.
\label{Yfcn}
\eeqn

Solving \eqref{Eqf}, we get 
 \be\label{fFSC}
f=\frac{2  }{\sqrt{4 b_{-1} b_1-b_0^2}} \arctan \left(\frac{\left(b_{-1}-b_0+b_1\right)
   \tanh \left(\frac{\pi\phi
   }{\beta_\phi}\right)+b_{-1}-b_1}{\sqrt{4 b_{-1}
   b_1-b_0^2}}\right) +c_1\,.
\ee
We set $c_1=0$, since $c_1$ shift the origin of $\tilde{\phi}$. Also we set $b_1=b_{-1}$ such that the center point of the interval is at $\phi=0$. Furthermore, we define
\be\label{bs}
b_0=\frac{\pi}{\tilde\beta_\phi} \Big( \tanh(\frac{\pi l_\phi  }{2\beta_\phi}) +\coth(\frac{\pi l_\phi }{2\beta_\phi})   \Big), \quad
b_1=-\frac{\pi}{\tilde\beta_\phi} \csch(\frac{\pi l_\phi  }{\beta_\phi})\,,
\ee
then, the solution  \eqref{fFSC} can be written as
\be
f=\frac{\tilde\beta_\phi}{\pi}\arctanh \Big(  \frac{\tanh(\frac{\pi \phi  }{\beta_\phi}) }{\tanh(\frac{\pi l_\phi  } {2\beta_\phi})}   \Big)\,,
\ee

Solving \eqref{Eqg} with $Y$ given in \eqref{Yfcn}, we can get the solution of $g$. Similarly we fix the translation symmetry of $\tilde{u}$ and set the center points of the interval to be at $u=0$.  Furthermore we define the new parameters  $\tilde \beta_u , \tilde\beta_\phi$ and express $d_1,d_0,d_{-1}  $ as
\besub\label{ds}
\beqn
 d_1=d_{-1}=-\frac{\pi\csch(\pi l_\phi/\beta_\phi) \Big[ - \tilde\beta_u \beta_\phi^2+\pi  (l_\phi \beta_u+l_u\beta_\phi)  \tilde\beta_\phi \coth(\pi l_\phi /\beta_\phi )\Big] }{\beta_\phi^2\tilde \beta_\phi^2}\,,
 \\
 d_0= \frac{\pi\csch^2(\pi l_\phi/\beta_\phi) \Big[  - \tilde\beta_u \beta_\phi^2 \sinh(2\pi l_\phi/\beta_\phi)
 +2\pi  (l_\phi \beta_u+l_u\beta_\phi)  \tilde\beta_\phi \Big] }{\beta_\phi^2\tilde \beta_\phi^2}\,,
\eeqn
\eesub
 then the solution becomes 
\be
g=\frac{\tilde \beta_\phi }{2\beta_\phi^2}  \frac{2\beta_u\sinh(\pi l_\phi/\beta_\phi)-(l_\phi \beta_u+l_u\beta_\phi)\sinh(2\pi\phi/\beta_\phi)}
{ \cosh(\pi l_\phi/\beta_\phi)-\cosh(2\pi \phi/\beta_\phi)}  
-\frac{\tilde \beta_u}{\tilde \beta_\phi} f\,.
\ee
Finally, we obtain the Rindler transformation for BMSFT at finite temperature
\besub\label{rindlertFSC}
\beqn
\tilde \phi &=& \frac{\tilde\beta_\phi}{\pi}\arctanh \Big(  \frac{\tanh(\frac{\pi \phi  }{\beta_\phi}) }{\tanh(\frac{\pi l_\phi  } {2\beta_\phi})}   \Big)\,,
\\
\tilde u +\frac{\tilde \beta_u}{\tilde \beta_\phi}  \tilde \phi &=  & \frac{\tilde\beta_\phi}{\beta_\phi} 
\frac{\sinh(\pi l_\phi/\beta_\phi)}{ \cosh(\pi l_\phi/\beta_\phi)-\cosh(2\pi \phi/\beta_\phi)}  
\nonumber \\  &&
 \times \Big(
u+\frac{\beta_u}{\beta_\phi} \phi   -\frac{l_\phi \beta_u+l_u\beta_\phi}{2\beta_\phi}\csch(\pi l_\phi/\beta_\phi) \sinh(2\pi \phi /\beta_\phi)
\Big)\,.
\eeqn
\eesub

With the Rindler transformation coefficients \eqref{ds} and \eqref{bs} known, one can also write down the modular flow and its bulk extension using \eqref{dtphi}, \eqref{dtu}, \eqref{ktLM} and the Killing vectors \eqref{KillingFSC}.

\subsection{BMSFT on the cylinder} \label{MinkTs}

The BMSFT with a spatial circle $\phi\sim\phi+2\pi$, can be obtained from thermal BMSFT through analytical continuation by setting $\beta_\phi=-i 2 \pi, \beta_u=0$,
\besub\label{rindlertGM}
\beqn
\tilde \phi &=& \frac{\tilde\beta_\phi}{\pi}\arctanh \frac{\tan(\frac{\phi}{2})}{\tan (\frac{l_\phi}{4})}\,,
\\
\tilde u +\frac{\tilde \beta_u}{\tilde \beta_\phi}  \tilde \phi &=  & \frac{\tilde \beta_\phi \sin(l_\phi/2)}{2\pi( \cos \phi-\cos (l_\phi/2))}
\Big( u-\frac{1}{2} l_u \csc(\frac{l_\phi}{2}) \sin \phi  \Big)\,.
\eeqn
\eesub
The modular flow is given by
\begin{align}\label{ktGMB}
k_t=&\pi  \csc \left(\frac{l_{\phi }}{2}\right) \left(-l_u \csc \left(\frac{l_{\phi }}{2}\right)+l_u \cos (\phi ) \cot \left(\frac{l_{\phi }}{2}\right)+2 u \sin (\phi )\right)\partial_u
\cr
&+2 \pi  \left(\cos \left(\frac{l_{\phi }}{2}\right)-\cos (\phi )\right) \csc \left(\frac{l_{\phi }}{2}\right)\partial_\phi\,,
\end{align}
which vanishes at the boundary end points $(-\frac{l_u}{2},-\frac{l_\phi}{2})$ and $(\frac{l_u}{2},\frac{l_\phi}{2})$. Substitute \eqref{KillingGM} into \eqref{dtphi} and \eqref{dtu},  the bulk extension of the modular flow in global Minkowski is given by
%
\begin{align}\label{ktGM}
k_t^{\text{bulk}} =&\pi  \csc \left(\frac{l_{\phi }}{2}\right) \left(-l_u \csc \left(\frac{l_{\phi }}{2}\right)+l_u \cos (\phi ) \cot \left(\frac{l_{\phi }}{2}\right)+2 u \sin (\phi )\right)\partial_u
\cr
 &+\frac{\pi  \left(\cot \left(\frac{l_{\phi }}{2}\right) \left(l_u \sin (\phi ) \csc \left(\frac{l_{\phi }}{2}\right)+2 r\right)-2 (r+u) \cos (\phi ) \csc \left(\frac{l_{\phi }}{2}\right)\right)}{r}\partial_\phi
 \cr
 &-\pi  \csc \left(\frac{l_{\phi }}{2}\right) \left( l_u \cos (\phi ) \cot \left(\frac{l_{\phi }}{2}\right) + 2 (r+u) \sin (\phi )\right)\partial_r\,.
\end{align}



\bibliographystyle{JHEP}
 \bibliography{EEflat}

 \end{document}

%% file: ShortCommand.tex
\newcommand{\be}{\begin{equation}}
\newcommand{\ee}{\end{equation}}
\newcommand{\bpm}{\begin{pmatrix}}
\newcommand{\epm}{\end{pmatrix}}

\newcommand{\beqn}{\begin{eqnarray}}
\newcommand{\eeqn}{\end{eqnarray}}

\newcommand{\Lie}{\mathcal{L}}

\newcommand{\besub}{\begin{subequations}}
\newcommand{\eesub}{\end{subequations}}
\newcommand{\bea}{\begin{eqnarray}}
\newcommand{\eea}{\end{eqnarray}}
\newcommand{\p}{\partial}
\def\tp{\tilde{\partial}}

\DeclareMathOperator{\csch}{csch}
\DeclareMathOperator{\arctanh}{arctanh}

\DeclareMathOperator{\Tr}{Tr}